\documentclass[fleqn,usenatbib]{mnras}

\usepackage{newtxtext,newtxmath}

\usepackage[T1]{fontenc}
\usepackage{ae,aecompl}

\DeclareRobustCommand{\VAN}[3]{#2}
\let\VANthebibliography\thebibliography
\def\thebibliography{\DeclareRobustCommand{\VAN}[3]{##3}\VANthebibliography}

\usepackage{gensymb}

\usepackage{graphicx}	
\usepackage{amsmath}	
\usepackage{ulem}
\usepackage{color}     
\usepackage{svg}
\usepackage{enumitem}
\renewcommand{\vec}[1]{\mathbf{#1}}
\newcommand{\hvec}[1]{\hat{\mathbf{#1}}}
\newcommand{\DS}{\displaystyle}
\newcommand{\pd}[2]{\frac{\partial #1}{\partial #2} }
\newcommand{\HALF}{\frac{1}{2}}

\newcommand{\av}[1]{\left< {#1} \right>}
\newcommand{\quotes}[1]{``#1''}
\newcommand{\tens}[1]{\mathsf{#1}}

\newcommand{\cE}{\mathcal{E}}
\newcommand{\cF}{\mathcal{F}}
\newcommand{\cH}{\mathcal{H}}

\newcommand{\cR}{\mathcal{R}}

\newcommand{\cW}{\mathcal{W}}

\newcommand{\cc}{ {\boldsymbol{c}} }

\newcommand{\xf}{ {\mathbf{x}_f} }
\newcommand{\yf}{ {\mathbf{y}_f} }
\newcommand{\zf}{ {\mathbf{z}_f} }

\newcommand{\xe}{ {\boldsymbol{x}_e} }
\newcommand{\ye}{ {\boldsymbol{y}_e} }
\newcommand{\ze}{ {\boldsymbol{z}_e} }

\DeclareMathOperator{\sech}{sech}

\usepackage{newtxtext,newtxmath}

\title[Fourth-order method for Res-RMHD]{A fourth-order accurate finite volume scheme for resistive relativistic MHD}

\author[Mignone et al.]{
A. Mignone$^{1},$\thanks{E-mail: andrea.mignone@unito.it}
V. Berta$^{1}$,
M. Rossazza$^{1}$,
M. Bugli$^{1,2,3}$,
G. Mattia$^{4}$,
L. Del Zanna$^{5,4,6}$, and
L. Pareschi$^{7,8}$
\\
$^{1}$Dipartimento di Fisica, Università degli Studi di Torino , Via Pietro Giuria 1, I-10125 Torino, Italy\\
$^{2}$INFN - sezione di Torino , Via Pietro Giuria 1, I-10125 Torino, Italy\\
$^{3}$Université Paris-Saclay, Université Paris Cité, CEA, CNRS, AIM, F-91191, Gif-sur-Yvette, France\\
$^{4}$INFN - sezione di Firenze, Via G. Sansone 1, I-50019 Sesto Fiorentino (FI), Italy \\
$^{5}$Dipartimento di Fisica e Astronomia, Università degli Studi di Firenze, Via G. Sansone 1, I-50019 Sesto Fiorentino (FI), Italy \\
$^{6}$ INAF - Osservatorio Astrofisico di Arcetri, L.go E. Fermi 5, I-50125 Firenze, Italy \\
$^{7}$ Maxwell Institute for Mathematical Sciences and
Department of Mathematics, Heriot-Watt University, Colin Maclaurin Building, Edinburgh, EH14 4AS, United Kingdom\\
$^{8}$ Dipartimento di Matematica e Informatica, Università degli Studi di Ferrara, Via N. Machiavelli 30, I-44121 Ferrara, Italy
}
\date{Accepted XXX. Received YYY; in original form ZZZ}

\pubyear{2023}

\begin{document}

\maketitle


\begin{abstract}
We present a finite-volume, genuinely $4^{\rm th}$-order accurate numerical method for solving the equations of resistive relativistic magnetohydrodynamics (Res-RMHD) in Cartesian coordinates.
In our formulation, the magnetic field is evolved in time in terms of face-average values via the constrained-transport method while the remaining variables (density, momentum, energy and electric fields) are advanced as cell volume-averages.  
Spatial accuracy employs $5^{\rm th}$-order accurate WENO-Z reconstruction from point values (as described in a companion paper) to obtain left and right states at zone interfaces.
Explicit flux evaluation is carried out by solving a Riemann problem at cell interfaces, using the Maxwell-Harten-Lax-van Leer with contact wave resolution (MHLLC).
Time stepping is based on the implicit-explicit (IMEX) Runge-Kutta (RK) methods, of which we consider both the $3^{\rm rd}$-order strong stability preserving SSP3(4,3,3) and a recent $4^{\rm th}$-order additive RK scheme, to cope with the stiffness introduced by the source term in Ampere's law. 
Numerical benchmarks are presented in order to assess the accuracy and robustness of our implementation.
\end{abstract}



\section{Introduction}
%
%

The investigation of relativistic plasma dynamics holds crucial importance in understanding the complex nature of high-energy astrophysical phenomena. 
Under typical astrophysical conditions, resistivity remains exceptionally low, and the ideal framework well describes processes occurring on rapid time scales. 
Nevertheless, the evolving flow dynamics can cause the formation of local regions with steep gradients, such as current sheets, where the influence of resistivity can no longer be ignored.

It appears then clear the necessity of numerical simulations of astrophysical plasmas beyond the ideal magnetohydrodynamics (MHD) assumptions.
Resistive Relativistic Magnetohydrodynamics (Res-RMHD) is indeed a framework that takes into account the contribution of finite plasma conductivity. 
In Res-RMHD, the comoving electric field no longer vanishes as in ideal MHD and it is directly proportional to the local current density, which is characterized by a scalar resistivity coefficient denoted as $\eta$.

Numerical methods for the solutions of the Res-RMHD equations have been now presented by a number of authors in the context of $2^{\rm nd}$-order conservative schemes \citep[see, for instance,][and references therein]{Palenzuela_etal2009, Komissarov_2007, Bucciantini_delZanna2013, Dionys_etal2013, Mizuno2013, Miranda_etal2018, Mignone_etal_2019, Cheong_etal2022, Nakamura_2023} or higher order schemes \citep{Dumbser_Zanotti2009, DelZanna_etal_2014, Bugli_etal2014, Tomei2020}.

A delicate aspect for their robust implementation is the stability of the method while dealing with the stiff relaxation term in Ampere's law, posing very strict constraints for time-explicit calculations. 
This makes the implementation of Res-RMHD numerical schemes more challenging than their ideal counterpart.
In this sense, the employment of implicit–explicit Runge–Kutta (IMEX, see \citealt{Pareschi_Russo2005}) represents an effective solution to the problem, allowing larger time steps to overcome stiffness. 

Another important aspect is related to the divergence-free condition and charge conservation which are not automatically fulfilled at the discrete level.
To cope with this issue, here we follow an approach similar to \cite{Mignone_etal_2019} (paper I henceforth), by adopting the constrained transport (CT) approach to control the divergence of magnetic field. 
In the CT formalism \citep[see][and references therein]{Evans_Hawley1988, Balsara_Spicer1999, Londrillo_DelZanna2004, Mignone_DelZanna2021} the magnetic field components are evolved as area-weighted quantities positioned on the faces orthogonal to the corresponding component direction.
This ensures that the condition $\nabla\cdot\nabla\times\vec{E}=0$ is satisfied also at discrete level.
In paper I, the CT method has been extended to control the divergence of the electric field as well, thus ensuring a consistent approach for charge conservation.
The resulting scheme has shown to be more robust when compared to cell-centered frameworks such as  the generalized Lagrange multipliers \citep[GLM, ][]{Munz_etal2000, Dedner_etal2002}.
In the present work, however, we choose to constrain only the magnetic field and evolve the electric field as a zone-centered variable without explicitly enforcing its divergence, simply replacing the charge density with $\nabla\cdot\vec{E}$ where needed. 
As demonstrated later and from extensive testing (not part of this work), we have not found any appreciable evidence of loss of accuracy or stability. 
Similar conclusions have been reported by \cite{Bucciantini_delZanna2013,DelZanna_etal_2016, Tomei2020}.

While $2^{\rm nd}$-order numerical schemes already provide an effective tool for solving such equations, higher-order methods can play a crucial role in the context of Res-RMHD simulations. 
An increase in overall accuracy and convergence order of the numerical integration, in fact, can significantly decrease the numerical dissipation associated to the scheme's discretization errors.
The benefits of such improvements are twofold.
On the one hand, higher-order schemes can generally lead to higher computational efficiencies, thus decreasing the computational time required to achieve a given accuracy \citep{Berta_etal_2024}.
On the other hand, a decrease in numerical dissipation means that we can probe the dynamical impact of more realistic (i.e. lower) values of physical magnetic dissipation, which needs to dominate over the numerical one in any reliable Res-RMHD model \citep{Mattia2023}.

In this paper we extend the $4^{\rm th}$-order accurate finite volume (FV) method described in \citet{Berta_etal_2024} (paper II henceforth) to the Res-RMHD equations.
Our scheme employs pointwise WENO-Z spatial reconstruction \citep[see paper II and the original paper by][]{Borges_WENOZ2008} and either the strong stability preserving SSP3(4,3,3) IMEX Runge–Kutta of \cite{Pareschi_Russo2003, Pareschi_Russo2005} or the more recent IMEX additive Runge-Kutta scheme ARK4(3)7L[2]SA$_1$ by \cite{Kennedy_Carpenter2019} to achieve, respectively, $3^{\rm rd}$- or $4^{\rm th}$-order accuracy in time. 
We refer to \cite{Carpenter2005, Conde2017, Boscheri2021} for other IMEX schemes of high order. Note that, the use of pointwise spatial reconstructions is indeed crucial to achieve accuracy beyond second order when evaluating the implicit stiff source terms within the IMEX formulation, as discussed recently in \cite{Boscarino2018}.

The rest of the paper is structured as follows. 
In \S\ref{sec:equations}, we review the fundamental equations of Res-RMHD framework. 
In \S\ref{sec:num_method} our $4^{\rm th}$-order FV-CT method is illustrated and in \S\ref{sec:results} we assess the robustness of the algorithm through extensive numerical benchmarking.
Results are summarized in \S\ref{sec:summary}.
\section{Resistive relativistic MHD equations}
\label{sec:equations}
%

When dealing with relativistic plasmas, it is convenient to start from the appropriate set of covariant equations. In the (single fluid) MHD approximation, these consist of the Maxwell equations
\begin{equation}
  \nabla_\mu F^{\mu\nu}  = -J^\nu , \qquad\qquad
  \nabla_\mu F^{*\mu\nu} = 0,
\end{equation}
where $F^{\mu\nu}$, $F^{*\mu\nu}$, $J^\mu$ are, respectively, the Faraday tensor, its dual, the four-current density, and in the conservation laws for mass and total (gas and electromagnetic fields) energy-momentum
\begin{equation}\label{eq:cons}
  \nabla_\mu(\rho u^\mu) = 0 ,\qquad\qquad \nabla_\mu T^{\mu\nu} \equiv \nabla_\mu (T^{\mu\nu}_\mathrm{gas} + T^{\mu\nu}_\mathrm{em}) = 0,
\end{equation}
where $\rho$ is the rest mass density, $u^\mu$ is the four-velocity of the fluid, and $T^{\mu\nu}$ is the total energy-momentum tensor. 
The two contributions are, respectively
\begin{equation}
\begin{array}{lcl}
T^{\mu\nu}_\mathrm{gas}    & = & \rho h \, u^\mu u^\nu + pg^{\mu\nu}, \\ \noalign{\medskip}
T^{\mu\nu}_\mathrm{em} & = &  F^\mu_{\,\,\,\lambda}F^{\nu\lambda}  - \tfrac{1}{4}(F_{\lambda\kappa}F^{\lambda\kappa}) g^{\mu\nu},
\end{array}
\end{equation}
where $h$, $\varepsilon$, $p$ are respectively the gas specific enthalpy, energy density and pressure, as measured in the fluid rest frame and related by $\rho h = \varepsilon + p$. These thermodynamic quantities must be also connected by an \textit{Equation of State} (EoS), for instance, given in the form $h=h(\rho,p)$.

It is convenient to express the Faraday tensor and its dual in terms of the four-velocity and the magnetic $b^\mu$ and electric $e^\mu$ fields measured in the fluid rest frame
\begin{equation}
\begin{array}{lcl}
F^{\mu\nu}  & = & u^\mu e^\nu - u^\nu e^\mu + \epsilon^{\mu\nu\lambda\kappa}b_\lambda u_\kappa, \\
F^{*\mu\nu} & = & u^\mu b^\nu - u^\nu b^\mu - \epsilon^{\mu\nu\lambda\kappa}e_\lambda u_\kappa,
\end{array}
\end{equation}
where $\epsilon^{\mu\nu\lambda\kappa}$ represents the Levi-Civita pseudo-tensor.
Ideal relativistic MHD is obtained by assuming a vanishing electric field in the frame comoving with the fluid, while in the resistive case we will assume a simple Ohmic relation
\begin{equation}\label{eq:ohm}
e^\mu = \eta j^\mu,
\end{equation} 
where $j^\mu = J^\mu + (J^\nu u_\nu) u^\mu$ is the current density measured in the fluid frame and $\eta$ is the resistivity coefficient, supposed to be a scalar for the sake of simplicity. 

Assuming from now on a Minkowski flat spacetime and splitting time and space, it is convenient to write the fluid velocity as
\begin{equation}
 u^\mu = (\sqrt{1 + u^2},\vec{u}) \equiv (\gamma,\gamma\vec{v}),
\end{equation}
where $\vec{v}$ is the usual three-dimensional velocity of the fluid measured in the laboratory frame and $\gamma$ the associated Lorentz factor, while the electromagnetic fields can be split as
\begin{equation}\label{eq:emu+bmu}
\begin{array}{lll}
e^\mu & = & (\vec{u}\cdot\vec{E}, \,\gamma \, \vec{E} + \vec{u}\times\vec{B}), \\ \noalign{\medskip}
b^\mu & = & (\vec{u}\cdot\vec{B}, \, \gamma \, \vec{B} - \vec{u}\times\vec{E}),
\end{array}
\end{equation}
where $\vec{E}$ and $\vec{B}$ are the usual electric and magnetic fields measured in the laboratory frame.
The four-current can be also split
\begin{equation}\label{eq:current}
J^\mu = (q,\vec{J}) = (q \, , \, q\vec{v} + \eta^{-1}\Tilde{\vec{E}}),
\end{equation}
in which $q$ is the charge density measured in the laboratory frame and $\vec{J}$ the spatial current density, with
\begin{equation}\label{eq:Etilde}
\Tilde{\vec{E}} \equiv \gamma\vec{E} + \vec{u}\times\vec{B} - (\vec{E}\cdot\vec{u})\vec{v} \Rightarrow \Tilde{\vec{E}} = \eta (\vec{J}- q\vec{v}),
\end{equation}
where Eqns. (\ref{eq:ohm}) and (\ref{eq:emu+bmu})  has been used.
Given the above spatial vectors, Maxwell's system is written in the usual way, a couple of evolutionary equations
\begin{equation}\label{eq:Maxwell}
  \pd{\vec{B}}{t} + \nabla\times\vec{E} = 0 , \qquad\qquad
  \pd{\vec{E}}{t} - \nabla\times\vec{B} = - \vec{J},
\end{equation}
and a couple of non-evolutionary constraints
\begin{equation}\label{eq:Maxwell_constraints}
  \nabla\cdot\vec{B} = 0 , \qquad\qquad \nabla\cdot\vec{E} = q,
\end{equation}
and we use the latter, Gauss's law, to define the charge density $q$.

The final system of resistive relativistic magnetohydrodynamics (RRMHD) is retrieved by splitting the covariant equations (\ref{eq:cons}), and by the first couple of Maxwell's equations (\ref{eq:Maxwell}). 
We obtain 11 evolution equations which may be written in the form of a set of conservation laws with source terms
\begin{equation}\label{eq:cons_laws}
 \pd{U}{t} = -\nabla\cdot\tens{F}(U) + S_e (U) + S(U),
\end{equation}
where 
\begin{equation}\label{eq:fluxTensor}
U = \left( \begin{array}{c}
D \\
m_i \\
{\cal E} \\
B_i \\
E_i 
\end{array}\right), \quad
 \tens{F}_j = \left( \begin{array}{c}
  \rho u_j \\
  \rho h u_i u_j  -E_iE_j -B_iB_j + p_\mathrm{tot}\delta_{ij} \\
   m_j \\
   \varepsilon^{ijk} E_k \\
  -\varepsilon^{ijk} B_k
 \end{array}\right) ,
\end{equation}
with $\varepsilon^{ijk}$ is the three-dimensional Levi-Civita symbol. Here the mass density $D=\rho\gamma$, the total momentum density $\vec{m}= D h\,\vec{u} + \vec{E}\times\vec{B}$, and the total energy density ${\cal E}=D h\gamma - p + u_\mathrm{em}$ (all measured in the laboratory frame) are the so-called \textit{conserved} variables, where $u_\mathrm{em} = (E^2 + B^2)/2$ is the electromagnetic energy density, and $p_\mathrm{tot}=p+u_\mathrm{em}$ is the total pressure. These variables are expressed in terms of the \textit{primitive} variables $\rho$, $\vec{v}$, and $p$.
Notice that the electric and magnetic fields $\vec{E}$ and $\vec{B}$ act as both conserved and primitive variables.

The source term is non-zero only in the equation for $\vec{E}$ and it contains the spatial current density $\vec{J}$ in Eq. (\ref{eq:current}). This is split, for computational purposes, into standard and \textit{stiff} ($\propto 1/\eta$) contributions, so that the source terms $S_e$ and $S$ in Eq. (\ref{eq:cons_laws}) are given by
\begin{equation} \label{eq:sources}
 S_e =  \left(\begin{array}{c}
          0_{\times8} \\
          - q v_i
        \end{array}\right)
 \,,\quad
  S = \frac{1}{\eta} \left( \begin{array}{c}
    0_{\times8} \\
    - \Tilde{E}_i
   \end{array}\right) ,
\end{equation}
where $\Tilde{\vec{E}}$ is defined in Eq. (\ref{eq:Etilde}).
Numerical stiffness is obviously given by the fact that the resistivity coefficient $\eta$ is negligible in real astrophysical plasmas, and finite but very small in computational applications, so the term $S$ can be very large and special care should be adopted for the evolution of the electric field (the terms $R$ and $S$ will be treated differently in the numerical time evolution of $U$). 

We recall that in ideal MHD, that is assuming the limit $\eta\to 0$, this latter equation is not evolved at all, the current does not need to be computed, and the electric field is a derived quantity, simply given by $\vec{E} = - \vec{v}\times\vec{B}$, so that $\Tilde{\vec{E}} = 0$.
\section{NUMERICAL METHOD}
\label{sec:num_method}
%
%

\subsection{Basic discretization}
We employ a Cartesian coordinate system with axes represented by the unit vectors $\hvec{e}_x = (1,0,0)$, $\hvec{e}_y =(0,1,0)$ and $\hvec{e}_z =(0,0,1)$, uniformly discretized into a regular mesh with coordinate spacing $\Delta x$, $\Delta y$ and $\Delta z$. 
We will frequently use the notation $\cc = (i,j,k)$ as a short-hand to identify a single computational cell centered at $(x_i, y_j, z_k)$ 
and delimited by the six interfaces orthogonal to the coordinate axis centered, respectively, at $(x_{i\pm\HALF},\, y_j,\, z_k)$, $(x_i, y_{j\pm\HALF}, z_k)$, and $(x_i, y_j, z_{k\pm\HALF})$.
The grid interfaces location will be short-handed with
%
\begin{equation}\label{eq:interface_notation}
  \begin{array}{l}
    \xf \equiv \left(i+\HALF, j, k\right),  \\ \noalign{\medskip}
    \yf \equiv \left(i, j+\HALF, k\right),  \\ \noalign{\medskip}
    \zf \equiv \left(i, j, k+\HALF\right) \,.
  \end{array}
\end{equation}
Likewise, we define the location of the cell edges as
%
\begin{equation}\label{eq:edge_notation}
  \begin{array}{l}
    \xe \equiv \left(i,j+\HALF,k+\HALF\right), \\ \noalign{\medskip}
    \ye \equiv \left(i+\HALF,j,k+\HALF\right), \\ \noalign{\medskip}
    \ze \equiv \left(i+\HALF,j+\HALF,k\right)\,.
  \end{array}
\end{equation}

Integrating the original PDE (Eq. \ref{eq:cons_laws}) over a cell volume for $U=\{D, \vec{m}, \cE, \vec{E} \}$ gives the semi-discrete form
\begin{equation}\label{eq:discrete_update}
  \frac{d\av{U}_{\cc}}{dt} = \hat{R}_\cc + \av{S
  }_\cc \,,
\end{equation}
where $\av{U}_{\cc}$ denotes the volume average of the corresponding conserved quantity, $\av{S}_\cc$ is the stiff part of the source term (second term in Eq. \ref{eq:sources}), while
\begin{equation}\label{eq:rhs}
  \arraycolsep=1.8pt\def\arraystretch{1.2}
  \begin{array}{lcl}
  \DS \hat{R}_\cc &=&\DS  -
  \left(\frac{  \hat{F}_{x, \xf} 
              - \hat{F}_{x, \xf - \hvec{e}_x} } 
             {\Delta x}\right)
              -\left(\frac{  \hat{F}_{y, \yf} 
                    - \hat{F}_{y, \yf - \hvec{e}_y} } 
             {\Delta y}\right)
  \\ \noalign{\medskip}           
  & &\DS -\left(\frac{  \hat{F}_{z, \zf} 
              - \hat{F}_{z, \zf - \hvec{e}_z} } 
             {\Delta z}\right) + \av{S_e}_\cc   \,,
  \end{array}
\end{equation}
is the explicit right-hand side including surface flux contributions plus the volume average of the explicit source term $S_e$ (first term in Eq. \ref{eq:sources}) whose only non-vanishing component is the $q\vec{v}$ term in the electric field update.

In Eq. (\ref{eq:rhs}), $\hat{F}_{x, \xf}$ (and likewise $\hat{F}_{y, \yf}$ and $\hat{F}_{z, \zf}$) represents the surface-integrated numerical flux across an $x$-face, i.e.:
\begin{equation}\label{eq:flux}
  \hat{F}_{x,\xf} 
   =
  \frac{1}{\Delta y\Delta z}
  \int F_{x,\xf} dy\,dz  \,,
\end{equation}
which has to be computed with a high-order quadrature rule.
Expressions for the $y$- and $z$-directions are obtained in a similar fashion.

Magnetic field variables are treated as staggered quantities and are evolved using the integral form of Faraday's law (the first in Eq. \ref{eq:Maxwell}) and direct application of Stokes' theorem (see \S 2, \S 3 and \S 4 of paper II for more details).
Thus, if $\hat{B}_{x,\xf}$, $\hat{B}_{y,\yf}$ and $\hat{B}_{z,\zf}$ are the surface-averaged components of $\vec{B}$ lying, respectively, on the $x$-, $y$- and $z$-interfaces, the constrained transport \citep[CT, see][]{Yee1966, Brecht_etal1981, Evans_Hawley1988, Balsara_Spicer1999} algorithm entails the following discrete update:
\begin{equation} \label{eq:stokes}
  \arraycolsep=1.8pt\def\arraystretch{1.2}
 \begin{array}{l}
  \DS \frac{d\hat{B}_{x,\xf}}{dt} = \DS - 
       \left(  \frac{\bar{E}_{z,\ze} - \bar{E}_{z,\ze-\hvec{e}_y}}
                    {\Delta y}
              -\frac{\bar{E}_{y,\ye} - \bar{E}_{y,\ye-\hvec{e}_z}}
                    {\Delta z}
       \right)  \,, \\ \noalign{\medskip}         
  \DS \frac{d\hat{B}_{y,\yf}}{dt} = \DS - 
       \left(  \frac{\bar{E}_{x,\xe} - \bar{E}_{x,\xe-\hvec{e}_z}}
                    {\Delta z}
              -\frac{\bar{E}_{z,\ze} - \bar{E}_{z,\ze-\hvec{e}_x}}
                    {\Delta x}
       \right)  \,, \\ \noalign{\medskip}         
  \DS \frac{d\hat{B}_{z,\zf}}{dt} = \DS - 
       \left(  \frac{\bar{E}_{y,\ye} - \bar{E}_{y,\ye-\hvec{e}_x}}
                    {\Delta x}
              -\frac{\bar{E}_{x,\xe} - \bar{E}_{x,\xe-\hvec{e}_y}}
                    {\Delta y}  
       \right)\,.
\end{array}
\end{equation}
In the previous expressions we have denoted the line-averaged electric field (electromotive force, or emf) at zone edges with $\bar{E}_{x,\xe}$, $\bar{E}_{y,\ye}$ and $\bar{E}_{z,\ze}$.
The actual computations of the emf is detailed in \S\ref{sec:CT_update}. 

Finally, it is worth to remind that in the presented $4^{\rm th}$-order scheme only magnetic field components are evolved as surface-average, while electric field (unlike paper I) is treated as a zone-centered variable rather than in a staggered fashion.

\subsection{Temporal update}

Equations (\ref{eq:discrete_update}) and (\ref{eq:stokes}) are evolved in time using implicit-explicit (IMEX) Runge Kutta (RK) schemes, for stiffly-accurate computations.
When applied to Eq. (\ref{eq:discrete_update}), an IMEX RK scheme can be expressed following the general formulation of \citet{Ascher1997, Pareschi_Russo2005}, yielding
\begin{equation}\label{eq:IMEX}
  \arraycolsep=1.8pt\def\arraystretch{1.2}
  \begin{array}{lcl}
    \av{U}_\cc^{(k)} &=& \DS \av{U}_\cc^n + \Delta t\sum_{j=1}^{k-1}\tilde{a}_{kj} \hat{R}_\cc
                        + \Delta t\sum_{j=1}^{k}{a}_{kj}\av{S}_\cc^{(j)} ,
    \\ \noalign{\medskip}
    \av{U}_\cc^{n+1} &=& \DS \av{U}_\cc^n + \Delta t\sum_{j=1}^{\nu}\tilde{w}_{j}\hat{R}_\cc^{(j)}
                        + \Delta t\sum_{j=1}^{\nu}{w}_{j}\av{S}_\cc^{(j)} ,
  \end{array}
\end{equation}
where $\nu$ is the number of integration stages, $\Tilde{A}$ is a $\nu\times\nu$ lower triangular matrix  ($\Tilde{a}_{ij} = 0$, $\forall j \geq i$) enclosing the coefficients of the explicit right-hand side $\hat{R}_\cc$ (Eq. \ref{eq:rhs} or the right-hand side of Eq. \ref{eq:stokes}), while $A=(a_{ij})$ is also a $\nu\times\nu$ matrix including the coefficients for the implicit part of the right-hand side, $S^{(j)}$. 
We recall that the use of a diagonally implicit Runge-Kutta (DIRK) solver guarantees an explicit evaluation of the non-stiff terms and implicit evaluation only for the diagonally stiff terms.

The final stage is always explicit and it is expressed in terms of the coefficient vectors $\Tilde{w} = (\Tilde{w}_1, ..., \Tilde{w}_\nu)^T$, and $w = (w_1, ..., w_\nu)^T$.
IMEX Runge-Kutta schemes are conveniently represented with a double tableau in the usual Butcher notation,
\begin{equation}
\qquad\qquad\qquad\qquad
\def\arraystretch{0.4}
\begin{array}{c|c}
\Tilde{c}  &  \Tilde{A}     \\ 
\hline                      \\
           &  \Tilde{w}^T
\end{array}
\qquad
\begin{array}{c|c}
c          &       A        \\ 
\hline                      \\
           &       w^T 
\end{array}
\end{equation}
where the coefficients $\Tilde{c}$ and $c$ refer to the treatment of non-autonomous systems and are defined as
\begin{equation}
\label{eq::csumIMEXcoeff}
\qquad\qquad\qquad\qquad
    \Tilde{c}_i = \sum_{j = 1}^{i-1} \Tilde{a}_{ij} \, ,
    \quad 
    c_i = \sum_{j = 1}^{i} a_{ij} \,.
\end{equation}

In this paper we consider two IMEX time-stepping methods: i) the $3^{\rm rd}$-order accurate IMEX Strong Stability Preserving SSP3(4,3,3) scheme of \citet{Pareschi_Russo2005} (with 4 implicit, and 3 explicit stages) and ii) the $4^{\rm th}$-order IMEX ARK4(3)7L[2]SA$_1$ of \citet{Kennedy_Carpenter2019} (ARK4 hereafter) which is L-stable, stiffly accurate (SA) with 7 stages.
The explicit coefficients of both methods can be found in Appendix \ref{app:butcher_tableau}.

From an algorithmic viewpoint, during the explicit-implicit update we found more convenient to separate the explicit contributions (acting on volume averages) from the implicit part which, in our formulation, will be operated on pointwise values.
For the generic $k$-th stage in Eq. (\ref{eq:IMEX}), we therefore act as follows:
\begin{equation}\label{eq:IMEX_SplitStage}
  \arraycolsep=1.8pt\def\arraystretch{1.2}
  \begin{array}{lcl}
    \av{U}_\cc^{(k*)} &=& \DS \av{U}_\cc^n 
      + \Delta t\sum_{j=1}^{k-1}\left(
              \tilde{a}_{kj} \hat{R}_\cc
            +  {a}_{kj}\av{S}_\cc^{(j)}\right) ,
    \\ \noalign{\medskip}
      U^{(k)}_\cc &=& U^{(k*)}_{\cc} + \Delta t a_{kk} S^{(k)}_\cc \,.
  \end{array}
\end{equation}
where the former step $(k*)$ is explicit while the latter $(k)$ is implicit. 
The algorithm is summarized in \S\ref{sec:alg_summary}.

\subsection{Point value recovery of primitive variables}
\label{sec:pointvalues}

Our algorithm starts with the zone-averaged value of the solution $\av{U}_{\cc}$ inside a cell and the face-averaged values of the staggered components of the magnetic field, $\hat{\vec{B}}$.
At the very first step, we require the point value of the solution at the cell-center. 
Following paper II, this operation may be carried out using the Laplacian operator,
\begin{equation}\label{eq:v2p}
    U_\cc = \left(1 - \frac{\Delta}{24}\right) \av{U}_\cc + O(h^4) \, ,
\end{equation}
where $\Delta$ is the Laplacian operator first introduced in the context of high-order methods by \cite{Corquodale_Colella2011}, i.e.
\begin{equation} \label{eq:laplacian}
  \Delta \av{U}_{\cc}  \equiv \left(\Delta^{x} + \Delta^{y} + 
     \Delta^{z}\right)\av{U}_\cc \, ,
\end{equation}
where, e.g.,
\begin{equation} \label{eq:laplacian1D}
  \Delta^{x} \av{U}_{\cc}  =
  \DS \left (\av{U}_{\cc - \hvec{e}_x} - 2\av{U}_{\cc} + 
             \av{U}_{\cc + \hvec{e}_x} \right) \, ,
\end{equation}
and similarly of the $y$- and $z$-directions.
Note that this step applies only to cell-centered quantities, i.e., lab density, momentum, energy and electric fields. 
For the magnetic field, instead, one has to first recover its point values at face centers using only the transverse Laplacian, so that
\begin{equation}\label{eq:v2p_B}
  \begin{array}{l}
  \DS B_{x,\xf} = \left(1 - \frac{\Delta^y+\Delta^z}{24}\right)\hat{B}_{x, \xf}   \,,
  \\ \noalign{\medskip}
  \DS B_{y,\yf} = \left(1 - \frac{\Delta^z+\Delta^x}{24}\right)\hat{B}_{y, \yf}   \,, 
  \\ \noalign{\medskip}
  \DS B_{z,\zf} = \left(1 - \frac{\Delta^x+\Delta^y}{24}\right)\hat{B}_{z, \zf}   \,.
  \end{array}
\end{equation}
Then, following \cite{Felker_Stone2018}, the pointwise value of the magnetic field at the cell center is obtained using an unlimited high-order interpolation, i.e.:
\begin{equation}\label{eq:Bcp}
 \begin{array}{l}
  \DS B_{x, c} = - \frac{1}{16}(B_{x,\xf+\hvec{e}_x} + B_{x,\xf-2\hvec{e}_x}) 
                + \frac{9}{16}(B_{x,\xf} + B_{x,\xf-\hvec{e}_x}) \, ,
  \\ \noalign{\medskip}
  \DS B_{y, c} = - \frac{1}{16}(B_{y,\yf+\hvec{e}_y} + B_{y,\yf-2\hvec{e}_y}) 
                + \frac{9}{16}(B_{y,\yf} + B_{y,\yf-\hvec{e}_y}) \, ,
  \\ \noalign{\medskip}
  \DS B_{z, c} = - \frac{1}{16}(B_{z,\zf+\hvec{e}_z} + B_{z,\zf-2\hvec{e}_z}) 
                + \frac{9}{16}(B_{z,\zf} + B_{z,\zf-\hvec{e}_z}) \, .
  \end{array}
\end{equation}
Now that all conservative variables are available at zone centers as point values, we convert them to primitive variables, i.e., $V_\cc = {\mathcal V}(U_\cc)$.
The conversion between conservative and primitive variables cannot be written in closed analytical form as it requires the inversion of a nonlinear function.
Differently from paper I, here we prefer to follow an  approach similar to that of \citet{Palenzuela_etal2009},  which consists of subtracting the electromagnetic contributions from momentum and energy densities and then resorting to a standard relativistic hydro inversion scheme to find the pressure. 
In our implementation, we employ the root solver of \cite{Mignone_etal_2005} \citep[see also \S 2 in][]{Mignone_Bodo_HLLC_2005} that can be extended to different equations of state.

Note that Eq. (\ref{eq:v2p}) may produce unphysical states in presence of sharp gradients and in regions of rapidly changing profiles.
A limiting strategy to identify troubled cells is indeed necessary before operating any conversion operation to maintain numerical stability.
We discuss our fallback approach in \S 3.9 of the present paper, while the reader may consult \S 3.4 of paper II for a full description.

\subsection{Implicit update of pointwise quantities}
\label{sec:implicit_update}

During the implicit part of our update (see the second equation in \ref{eq:IMEX_SplitStage}), we operate directly on the point value of the electric field vector:
\begin{equation}\label{eq:E(u)_v1}
  \vec{E} = \vec{E}^* - \frac{1}{\tilde{\eta}}\left[\gamma\vec{E} + \vec{u}\times\vec{B} - (\vec{E}\cdot\vec{u})\vec{v}\right]\,,
\end{equation}
where, by analogy Eq. (\ref{eq:IMEX_SplitStage}), we identify $\vec{E}^* \equiv \vec{E}^{(k*)}$ (available from the last explicit stage) and $\tilde{\eta}=\eta/(a_{kk}\Delta t)$ which is now a non-dimensional quantity.
The previous equation provides a relation between the electric field $\vec{E}$ and the four-velocity $\vec{u}$.
Using
\begin{equation}
\vec{E}\cdot\vec{u} = \frac{\tilde{\eta}\gamma}{1+\tilde{\eta}\gamma} (\vec{E}^*\cdot\vec{u}),
\end{equation}
we can re-write Eq. (\ref{eq:E(u)_v1}) as an explicit function $\vec{E}\equiv \vec{E}(\vec{u})$:
\begin{equation}\label{eq:E(u)_v2}
  (\tilde{\eta}+\gamma) \vec{E}(\vec{u})= 
  \tilde{\eta}\,\vec{E}^* - \vec{u}\times\vec{B} +
  \frac{\tilde{\eta}}{1 + \tilde{\eta}\gamma}(\vec{E}^*\cdot\vec{u})\vec{u} \, ,
\end{equation}
where $\gamma = \sqrt{1+u^2}$.

Note that the four-velocity is not a conservative variable and it must be derived from the set of conserved variables $U$.
Since momentum does not change during the implicit step,  we use the implicit relation
\begin{equation}
   \vec{f}(\vec{u}) = \vec{m} 
      - \Big(Dh(\vec{u})\vec{u}
      + \vec{E}(\vec{u})\times\vec{B}\Big) = \vec{0} \,.
\end{equation}
to derive $\vec{u}$, ad in paper I.
Note also that $D$ and $\vec{B}$ are held constant during this step.
Following \citet{Bucciantini_delZanna2013}, we adopt a Newton-Broyden root-finding method to derive the four-velocity (and the other primitive variables) from the set of conserved variables. 
Additionally, we refer the reader to \citet{Mattia2023} for a generalization of the Taub equation of state \citep{Mignone_etal_2005} and to \citet{Tomei2020} for a more comprehensive treatment of the dynamo term in Ohm's law along with for any spacetime metric in general relativity.
The iterative multi-dimensional Newton-Brodyen algorithm can be sketched as follows:
\begin{enumerate}
\setlength\itemsep{1.2em}

\item \label{item:NB_begcycle}
At the beginning of any step $\kappa=0,1,2,...$, use Eq. (\ref{eq:E(u)_v2}) to obtain $\vec{E}=\vec{E}(\vec{u})$.
When $\kappa = 0$, an initial guess for the four-velocity is needed: if $\Tilde{\eta} > 1$, employ the value provided at the current time level, otherwise reckon that obtained from the ideal RMHD equations.

\item
Given $\vec{u}^{(\kappa)}$ and $\vec{E}^{(\kappa)}$, compute $\vec{f}^{(\kappa)} \equiv \vec{f}(\vec{u}^{(\kappa)})$ from the conservation of the total momentum density
\begin{equation}\label{eq:NB_solve}
  \vec{f}^{(\kappa)} = \vec{m} - \Big[Dh(\vec{u})\vec{u}
                               + \vec{E}(\vec{u})\times\vec{B}\Big]^{(\kappa)} \, .
\end{equation}
Notice that the laboratory density, magnetic field, total energy and the conserved momentum remain constant during this iteration scheme.
Eq. (\ref{eq:NB_solve}) represents a set of three nonlinear equations to be solved for $\vec{u}^{(\kappa)}$. 
Thermodynamics quantities are also written as functions of the four-velocity; for an ideal gas law one finds
\begin{equation}
h(\vec{u}) = 1 + \Gamma_1 \frac{p\gamma}{D} \,,
\end{equation}
with $\Gamma_1=\Gamma/(\Gamma-1)$ ($\Gamma = 4/3$ for  relativistically hot plasmas), and, from the conservation of the total energy, the pressure is derived as
\begin{equation}
p(\vec{u}) = \frac{\mathcal{E} - D\gamma - (E^2 + B^2)/2}{\Gamma_1\gamma^2 - 1}\,.
\end{equation}

\item
Using the Jacobian $\tens{J}=\partial\vec{f}/\partial\vec{u}$, obtain an improved guess of the four-velocity through
\begin{equation}
 \vec{u}^{(\kappa+1)} =  \vec{u}^{(\kappa)}
                        - \left(\tens{J}^{(\kappa)}\right)^{-1}\vec{f}^{(\kappa)}\,,
\end{equation}
with the explicit form of the Jacobian being
\begin{equation}
{\tens J}_{ij} \equiv \pd{f_i}{u_j} = -Dh\,\delta_{ij} - D\,u_i\pd{h}{u_j} - \epsilon_{ilm}\pd{E_l}{u_j}B_m \, .
\end{equation}
The derivatives of $h$ and $\vec{E}$ are given by
\begin{equation}
D \pd{h}{u^j} = - \frac{\Gamma_1}{\gamma^2\Gamma_1 - 1}\left[(\gamma h D + p) v_j + \gamma E_i \pd{E_i}{u_j}\right] \, ,
\end{equation}
given the relation $\partial \gamma/\partial u_j = u_j/\gamma = v_j$, and, from Eq. (\ref{eq:E(u)_v2}), by
\begin{equation}
  \begin{array}{l}
  \DS (\tilde{\eta} + \gamma) \pd{E_i}{u_j} = 
   - E_i v_j - \epsilon_{ijk} B_k
  \\ \noalign{\medskip}
 \DS \quad + \frac{\tilde{\eta}}{1 + \tilde{\eta}\gamma} 
\left[u_i E^*_j + (\vec{E}^*\cdot\vec{u})
\left( \delta_{ij} -\frac{\tilde{\eta}}{1 + \tilde{\eta}\gamma} u_iv_j  \right) \right],
  \end{array}
\end{equation}
which can be also compared to the Jacobian presented in \citet{Tomei2020} (for a vanishing dynamo coefficient $\xi=0$).

\item
Exit the iteration cycle if the error is less than some prescribed accuracy $|\vec{f}^{(\kappa)}|<\epsilon$, otherwise go back to step \ref{item:NB_begcycle} and let $\kappa\to \kappa+1$. For standard applications, our algorithm converges in $2-5$ iterations given a relative tolerance of $10^{-11}$.

\end{enumerate}

The above procedure is to be implemented at the beginning of all IMEX substeps at cell centers using pointwise quantities.
Furthermore, it automatically operates the conversion from conservative to primitive variables, which are then readily available for reconstruction at cell interfaces and flux computation. 
For this reason, this procedure must be extended also to a layer of ghost zones (e.g., $3$ for a $5$-point stencil reconstruction) in order to avoid extra communication at the end of the implicit step.

\subsection{Point value reconstruction of primitive variables}
\label{sec:reconstruction}

Reconstruction at interfaces is operated directly on the function point values rather than on the one-dimensional average quantities as firstly introduced by paper II.
In the following paragraphs, we summarize the pointwise $5^{\rm th}$-order weighted essentially non-oscillatory scheme of Borges et al. \citep[WENO-Z, see][]{Borges_WENOZ2008}.

The interface value presented in Eq. (23) of paper II at $x=x_{i+\HALF}$ is computed as the convex combination of $3^{\rm rd}$-order accurate interface values built on the three possible three-point sub-stencils $\{i-2,i-1,i\}$, $\{i-1,i,i+1\}$, and $\{i,i+1,i+2\}$, i.e.
\begin{equation}\label{eq:wenoz}
\begin{array}{lcl}
    \DS V^L_{i+\HALF} & = & \DS
    \omega_0 \frac{3V_{i-2} - 10V_{i-1} + 15V_i}{8}  \\ \noalign{\medskip}
 & + & \DS \omega_1 \frac{-V_{i-1} +  6V_{i}   + 3V_{i+1}}{8} \\ \noalign{\medskip}
 & + & \DS \omega_2 \frac{3V_{i}   +  6V_{i+1} - V_{i+2}}{8} \, ,
\end{array}
\end{equation}
with $\{ V_{i\pm2}, V_{i\pm1}, V_{i} \}$ point values.
The weights $\omega_l$, for $l=\{0,1,2\}$, are defined as
\begin{equation}
    \omega_l = \frac{\alpha_l}{\sum_{m} \alpha_m}, \hspace{0.5cm} \alpha_l = d_l\left(1+ \frac{|\beta_0-\beta_2|}{\beta_l+\epsilon}\right) \, .
\end{equation}
Here $d_l = \{d_0 = 1/16, d_1 = 5/8, d_2 = 5/16 \}$ denotes the optimal weights that reckon the $5^{\rm th}$-order accurate approximation of Eq. (23) of paper II, $\epsilon = 10^{-40}$ is a small number avoiding division by zero and $\beta_l$ are the smoothness indicators (Eq. 26 of paper II).
A detailed description of pointwise reconstructions is presented in \S 3 of paper II.

\subsection{Riemann solver \& flux computation}
\label{sec:riemann}

The integrand in Eq. (\ref{eq:flux}) is approximated by means of a Riemann solver which, following the traditional formalism of FV Godunov schemes, returns a stable and consistent numerical flux as a function of L/R interface states obtained in the previous section.
We will employ the MHLLC solver described in paper I, which considers Maxwell's and gasdynamics equations decoupled during this step. 
This leads to i) a pair of outermost electromagnetic waves leading to jumps in $\vec{E}$, $\vec{B}$ (and therefore total momentum and energy) and ii) an inner wave fan including two hydrodynamics shocks enclosing a contact wave.

At the practical level, we first obtain the transverse components of $\vec{E}$ and $\vec{B}$ using Maxwell's jump conditions.
Specializing at an $x$-interface:
\begin{equation}\label{eq:resHLLC_EBtilde}
  \arraycolsep=1.8pt\def\arraystretch{1.2}
  \begin{array}{lcl}
    \tilde{B}_y &=&\DS \frac{B_{y,L} + B_{y,R}}{2} + \frac{E_{z,R} - E_{z,L}}{2}
  \\ \noalign{\medskip}
    \tilde{B}_z &=&\DS \frac{B_{z,L} + B_{z,R}}{2} - \frac{E_{y,R} - E_{y,L}}{2}
  \\ \noalign{\medskip}
    \tilde{E}_y &=&\DS \frac{E_{y,L} + E_{y,R}}{2} - \frac{B_{z,R} - B_{z,L}}{2}
  \\ \noalign{\medskip}
    \tilde{E}_z &=&\DS \frac{E_{z,L} + E_{z,R}}{2} + \frac{B_{y,R} - B_{y,L}}{2} \,.
  \end{array}
\end{equation}
Normal components ($B_x$ and $E_x$) remain continuous.

Next, we compute the hydrodynamics flux $\cH^*$ using the HLLC relativistic solver of \cite{Mignone_Bodo_HLLC_2005}:
\begin{equation}
  \cH^* = \left\{\begin{array}{lcl}
    \cH_L  & {\rm if} & \lambda_L > 0   \,,          
    \\ \noalign{\medskip}
    \cH_L + \lambda_L(\cW^*_L - \cW_L)
                   & {\rm if} & \lambda_L < 0 < \lambda^* \,,  
    \\ \noalign{\medskip}
    \cH_R + \lambda_R(\cW^*_R - \cW_R)
                   & {\rm if} & \lambda^* < 0 < \lambda_R \,, 
    \\ \noalign{\medskip}
    \cH_R  & {\rm if} & \lambda_R < 0   \,,
  \end{array}\right.
\end{equation}
where $\cH_S = (D,\, Q_kv_x + p\delta_{kx},\, Q_x)_S$, $S=L,R$ denotes the left or right states, $k=x,y,z$ is the spatial component, $\cW_S = (D,\, Q_k,\, \cE_h)_S$, $\cE_h = w\gamma^2-p$ is the gas energy, and $Q_k = w\gamma^2v_k$ is the $k$-th component of the gas momentum.
The wave speeds $\lambda_{L/R}$ are estimated from the fastest and slowest speeds as in \cite{Mignone_Bodo_HLLC_2005}.
The states in the star region enclosing the contact mode are obtained as
\begin{equation}
  \cW^*_{S} = \DS\frac{1}{\lambda_S-\lambda^*}
  \left(\begin{array}{l}
          D   (\lambda-v_x)  \\ \noalign{\medskip}
          Q_k (\lambda-v_x) + (p^* - p)\delta_{kx}  
          \\ \noalign{\medskip}
          E (\lambda - v_x) + p^*\lambda^* - pv_x   \\ \noalign{\medskip}
  \end{array}\right)_S
\end{equation}
for $S=L,R$ and the speed of the contact mode $\lambda^*_L = \lambda^*_R = \lambda^*$ is found from the negative branch of the quadratic equation
\begin{equation}\label{eq:HLLCquadratic}
    \cH_{[\cE]}^{\rm hll}\left(\lambda^*\right)^2
  - \left(\cW_{[\cE_h]}^{\rm hll} + \cH^{\rm hll}_{[Q_x]}\right)\lambda^*
  + \cW_{[Q_x]}^{\rm hll} = 0 \,.
\end{equation}
Here the subscript $[.]$ picks a specific component of the array, while $p^* = \cH^{\rm hll}_{[Q_x]} - \lambda^* \cH^{\rm hll}_{[\cE_h]}$ is the pressure in the star region, and the superscript ${\rm hll}$ indicates the HLL-average state (see Eq. 73 and 74 in paper I).
Putting all together, we retrieve the final flux:
\begin{equation}
  \cF_{x,\xf} = \left(\begin{array}{l}
   \cH^*  \\ \noalign{\medskip}
    0_{\times 6} \end{array}\right)
    +
  \left(\begin{array}{c}
    0                         
    \\ \noalign{\medskip}
    \hvec{e}_x\cdot\tilde{T}
    \\ \noalign{\medskip}
    \hvec{e}_x\cdot(\tilde{\vec{E}}\times\tilde{\vec{B}})
    \\ \noalign{\medskip}
     \hvec{e}_x\times\tilde{\vec{E}} \\ \noalign{\medskip}
    -\hvec{e}_x\times\tilde{\vec{B}}
    \end{array}\right) \,.
\end{equation}
A similar procedure is followed to calculate the $y$- and $z$-interface fluxes.

Finally, $4^{\rm th}$-order accuracy at zone interfaces is ensured by computing the surface-averaged flux through 
\begin{equation}\label{eq:avFlux}
  \hat{F}_{x,\xf} = \left(1 + \frac{\Delta^y + \Delta^z}{24}\right)
                    \cF_{x,\xf}  \,.
\end{equation}
where $\Delta^{y,z}$ have been defined in  Eq. (\ref{eq:laplacian}) and Eq. (\ref{eq:laplacian1D}). 
Note that the fluxes are used to update density, momentum, energy and electric fields. 
Magnetic fields, instead, are advanced in a separate step using the constrained transport formalism (see \S 3.7).

\subsection{Constrained Transport update}
\label{sec:CT_update}

\begin{figure}
  \centering
  \includegraphics[width=0.4\textwidth]{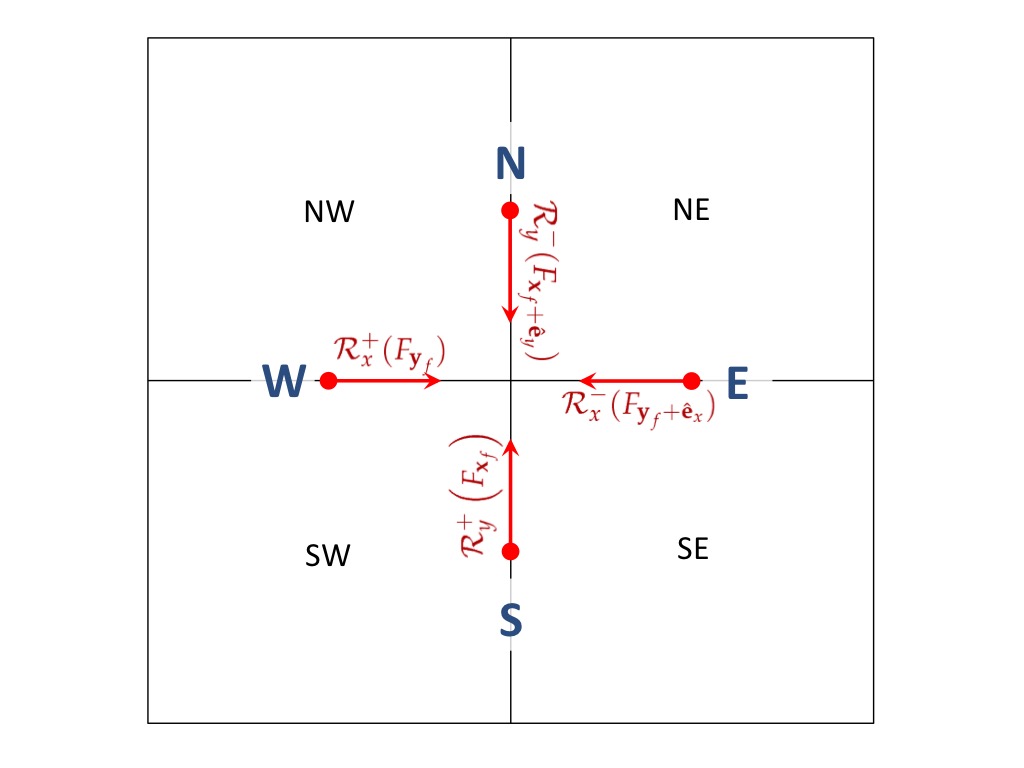}
  \caption{
   Top view of the intersection between four neighbor zones in the $xy$ plane.
   N, S, E, and W indicate the four cardinal directions with respect to the zone edge (here represented by the intersection between four neighbor zones), $\cR_x(F_{\yf})$ and $\cR_y(F_{\xf})$ are 1-D reconstruction operators applied to each zone face, see Eq. (\ref{eq:Erec}) and (\ref{eq:Brec}). }
  \label{fig:CT}
\end{figure} 

During the staggered update, we evolve the face-averaged components of magnetic fields using Eq. (\ref{eq:stokes}).
In order to retain $4^{\rm th}$-order accuracy we follow the steps already illustrated by paper II.
In what follows, we recap the procedure.

First, one needs to reconstruct the L/R transverse electric field components from the four faces sharing the same edge to that edge.
With reference to Fig. \ref{fig:CT}, which shows a top view of a $z$-edge in the $xy$ plane, this entails a reconstruction of the four adjacent values of $E_z$ (available at the $x$- and $y$-face centers) along the transverse directions ($y$ and $x$, respectively) to obtain four discontinuous values coming up at $\ze \equiv (i+\HALF, j+\HALF,k)$:
\begin{equation}\label{eq:Erec}
  \begin{array}{l}
  E^S_{z,\ze} = {\cal R}^+_y(E_{z,\xf}) \,,\quad
  E^N_{z,\ze} = {\cal R}^-_y(E_{z,\xf+\hvec{e}_y}) \,,\quad  
  \\ \noalign{\medskip}
  E^W_{z,\ze} = {\cal R}^+_x(E_{z,\yf}) \,,\quad
  E^E_{z,\ze} = {\cal R}^-_x(E_{z,\yf+\hvec{e}_x}) \,,\quad  
  \end{array}
\end{equation}
where the four cardinal directions in square brackets are referred to the edge location $\ze$.
For convenience, only one component per face has to be reconstructed\footnote{The approach followed here differs from the one employed in paper I, where the reconstruction is performed from the cell center using linear interpolation. 
Also, a typo is present in Eq. (48) of paper I, since the arithmetic average operator $\av{.}_z$ is not needed.}:
\begin{equation}
    E_{z,\xf} = \frac{E^L_{z,\xf} + E^R_{z,\xf}}{2} \,.
\end{equation}
In 3D, one has to repeat the reconstruction (Eq. \ref{eq:Erec}) for $E_{x,\yf}$ and $E_{x,\zf}$, respectively, along the $z$- and $y$-direction to $\xe$ and also for $E_{y,\xf}$ and $E_{y,\zf}$ along the $z$- and $x$-directions to $\ye$.

Next, we also reconstruct the face-centered point values of the normal components of the magnetic field towards the same edge:
\begin{equation}\label{eq:Brec}
  \begin{array}{l}
  B^S_{x,\ze} = {\cal R}^+_y(B_{x,\xf})   \,,   \quad
  B^N_{x,\ze} = {\cal R}^-_y(B_{x,\xf+\hvec{e}_y})      \,,
    \\ \noalign{\medskip}
  B^W_{y,\ze} = {\cal R}^+_y(B_{y,\yf})  \,,    \quad
  B^E_{y,\ze} = {\cal R}^-_y(B_{y,\yf+\hvec{e}_x}).
  \end{array}    
\end{equation}
Eqns (\ref{eq:Erec}) and (\ref{eq:Brec}) allow us to obtain the edge-centered point value of the upwind electric field, using the CT-Maxwell averaging scheme (see \S 3.4.2 of paper I):
\begin{equation}\label{eq:CT_Maxwell}
    E_{z,\ze} = \left[\frac{E^N_z + E^S_z + E^W_z + E^E_z}{4}
                + \frac{B^E_{y} - B^W_{y}}{2}
                - \frac{B^N_{x} - B^S_{y}}{2}\right]_\ze
\end{equation}
Similar expressions hold for $E_{x,\xe}$ and $E_{y,\ye}$ by suitable index permutation.

Finally, we obtain the electromotive force (that is, the averaged line-integral of the electric field needed in the discrete version of Faraday's law Eq. \ref{eq:stokes}) as
\begin{equation}\label{eq:emfz}
  \bar{E}_{z,\ze} = \left(1 + \frac{\Delta^z}{24}\right)E_{z,\ze} \,.
\end{equation}
Eqns. (\ref{eq:Brec})-(\ref{eq:emfz}) are extended to the 3D case by suitable index permutation.

\subsection{Explicit source term computation}
\label{sec:explicit_source}
In addition to the flux difference operator, the right-hand side (Eq. \ref{eq:rhs}) requires an explicit contribution from the $q\vec{v}$ source term during the electric field update.
To $4^{\rm th}$-order, this term should be included as a volume average over the zone, i.e., 
\begin{equation}
  \av{S}_e = \left(1 + \frac{\Delta}{24}\right)q_\cc\vec{v}_\cc \,,
\end{equation}
where $q_\cc$ and $\vec{v}_\cc$ are, respectively, the point values of the charge and the fluid velocity.
The former can be obtained from a cell-centered discretization of $\nabla\cdot\vec{E}$ yielding
\begin{equation}\label{eq:charge}
  q_\cc = \delta_xE_x + \delta_yE_y + \delta_zE_z  \,.
\end{equation}
Using the point value of the electric field already at disposal during the explicit step, we approximate (to $4^{\rm th}$-order) the $x$-contribution to the divergence as
\begin{equation}\label{eq:charge_dEx}
  \delta_x E_x = \frac{ 8\left(E_{x,\cc+\hvec{e}_x} - E_{x,\cc-\hvec{e}_x} \right) 
                       - \left(E_{x,\cc+2\hvec{e}_x} - E_{x,\cc-2\hvec{e}_x}\right)}{12\Delta x} \,.
\end{equation} 
Contributions along the $y$- and $z$-directions are obtained in a similar way.

We point out that the employment of unlimited differences in Eq. (\ref{eq:charge_dEx}) may lead to the appearance of spurious oscillations in the presence of strong gradients.
This does not represent an issue for the tests presented below, but it could be further improved by including information from the reconstructed interface values (limited).
We leave this investigation to a forthcoming work.

\subsection{Order reduction at discontinuities}
\label{sec:order_reduction}
As pointed out in paper II, $4^{\rm th}$-order accuracy cannot be maintained in proximity of discontinuous solutions or steep gradients where the function may not be differentiable. 
In order to detect such features we compute, at the beginning of the RK step, the derivative ratio sensor as outlined in Sec. 3.4 of paper II.
The detector flags computational cells where the order of the scheme can be locally reduced (this strategy is also known as \quotes{fallback} process) by i) not using the Laplacian during point value recovery and flux integration (Eq. \ref{eq:v2p} and \ref{eq:avFlux}) and ii) switching to either $3^{\rm rd}$-order WENO or $2^{\rm nd}$-order linear reconstruction.

The detector employs a five-zone stencil and does not require additional ghost zones.
A similar approach is also followed by \cite{Verma_etal2019}, albeit the authors make use of \quotes{global smoothness indicators}, defined in terms of weighted averages computed from the density and the three magnetic field components.

\subsection{Algorithm summary}
\label{sec:alg_summary}

To better outline our algorithm, let ${\cal D}$ be the set of zones comprising the active computational domain (e.g. no ghost zones).
Then ${\cal D}+1$ denotes the set of active computational zones augmented by one layer of ghost zones.
Since the last step of an IMEX-RK scheme is always explicit, our implementation can be best summarized starting from the end of the explicit stage $(k*)$ (first of Eq. \ref{eq:IMEX_SplitStage}) and moving on to the next, $k\to (k+1)*$:
\begin{enumerate}[align=left, label=\arabic*., leftmargin=*] 
  \item Start with the volume average of conservative variable, $\av{U}^{(k*)}_\cc$, defined on ${\cal D}$ at the end of the more recent explicit stage.
  \item Assign boundary conditions on $\av{U}_\cc$ in the first layer of ghost zones only so that $\av{U}_\cc$ is correctly specified on ${\cal D}+1$.
  \item Recover the point value of conservative ($U_\cc$) and primitive variables, $V_\cc = {\mathcal V}(U_\cc)$ (as shown in \S\ref{sec:pointvalues}) on ${\cal D}$ only.
  \item Carry out the implicit update by updating $V_\cc$ on ${\cal D}$ as outlined in \S\ref{sec:implicit_update}. 
  Notice that conservative variables $U_\cc$ do not change during this step, except for the electric field components.
  \item Assign boundary conditions on $V_\cc$. This will define primitive variables on ${\cal D}+n_g$, where $n_g = 3$.
  \item Using the point value of primitive variables, compute the implicit source term point value in ${\cal D}+1$, that is, $S_\cc = S(V_\cc)$.
  \item Compute the zone-averaged implicit source term $\av{S}_\cc$ on ${\cal D}$ using Simpson quadrature rule.
  \item Achieve the explicit step by constructing the right hand side $\hat{R}_\cc$ (Eq. \ref{eq:rhs}) and its staggered version (right-hand side of Eq. \ref{eq:stokes}): 

  \begin{itemize}
      \item Perform spatial reconstruction on primitive point value variables (\S\ref{sec:reconstruction});
      \item Solve a Riemann problems with L/R states at zone interfaces, see \S\ref{sec:riemann};
      \item Compute the surface-averaged flux using Eq. (\ref{eq:avFlux});
      \item Compute the point value of the electric field at zone edges (Eq. \ref{eq:CT_Maxwell}) following the procedure outlined in \S\ref{sec:CT_update};
      \item Obtain the line-averaged electro-motive force through Eq. (\ref{eq:emfz});
      \item Compute the volume-average of the explicit source term (\S\ref{sec:explicit_source}).
  \end{itemize}
  \item Complete the explicit step by obtaining the volume average conserved variables at the next RK stage, that is $\av{U}_\cc^{(k+1)*}$.
\end{enumerate}
%
%

Note that our formulation requires two boundary calls per stage: the first one is operated on $\av{U}_\cc$ but it extends only to one layer of ghost zones, while the second one specifies boundary values on the point value or primitive variables $V_\cc$ and it extends to $3$ ghost zones.
Unlike previous works, our formulation does not require more than $3$ ghost zones.

\section{Numerical Benchmarks}
\label{sec:results}
%
%

We now verify and assess the accuracy and robustness of our implementation through standard 2D and 3D reference solutions. 
Unless otherwise stated, we will employ the pointwise WENO-Z reconstruction (paper II) and the MHLLC Riemann solver coupled with the CT-Maxwell EMF average in all test problems. 
By default, order reduction (see Sec. \ref{sec:order_reduction}) is disabled unless otherwise stated.

Initial conditions are integrated with a 4 point Gaussian quadrature rule whenever smooth analytic functions are considered.

For quantitative purposes, we evaluate $L_1$ norm errors for a generic quantity $Q$ against a reference solution $Q_{\rm ref}$ as $\epsilon_1(Q) = \sum_i|Q(x_i)-Q_{\rm ref}(x_i)|/N$, where $x_i$ is a generic point of the domain and $N$ is the total number of grid zones.

All computations have been carried out by means of the PLUTO astrophysical code \citep{Mignone_PLUTO2007, Mignone_PLUTO2012}, where the $4^{\rm th}$-order resistive method has been implemented.
The Courant number is set to $0.4$ (in 2D) and to $0.3$ for $3D$ calculations.

\subsection{Telegraph Equation}

\begin{figure*}
    \centering
    \includegraphics[width=0.8\textwidth]{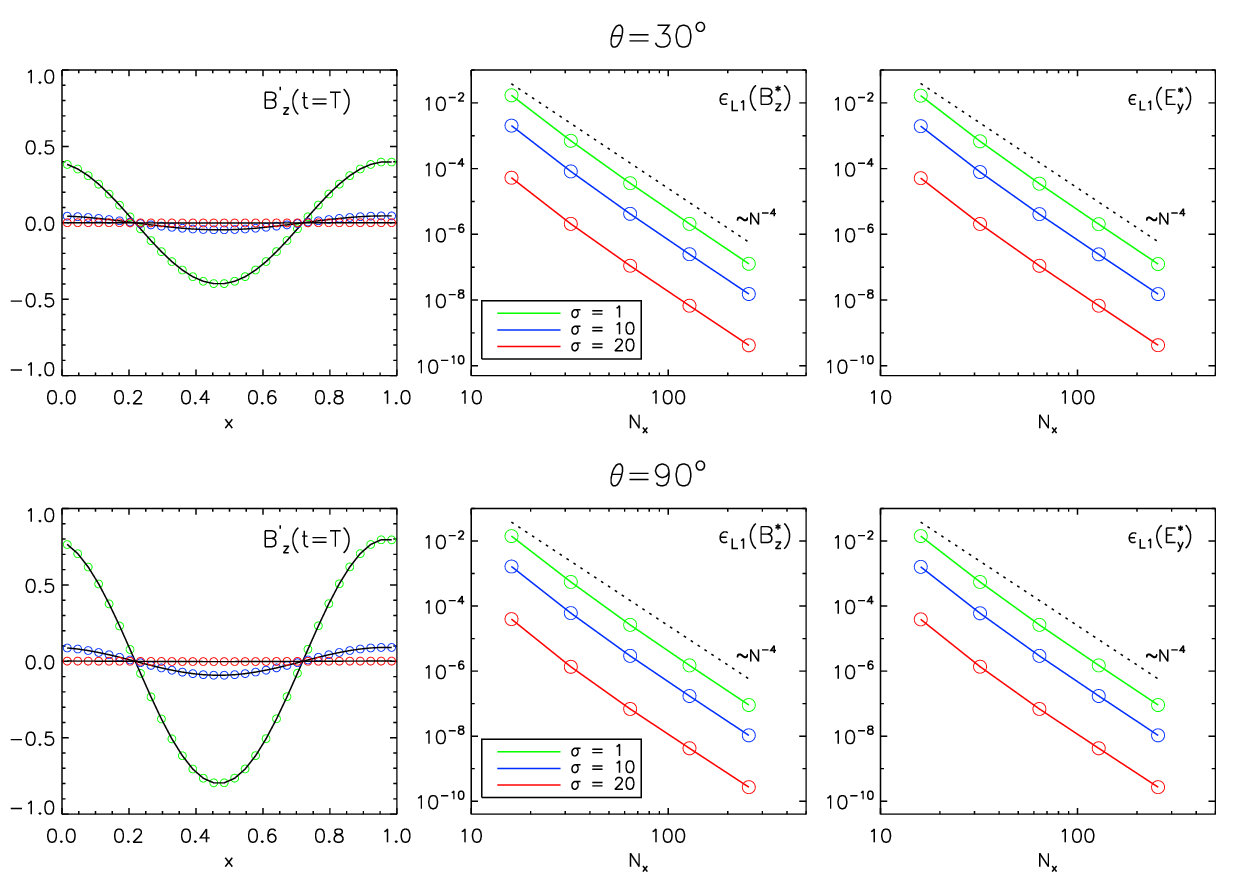}
    \caption{\small Numerical results of the telegraph equation test, for $\Theta = 30^\circ$ (top panels) and $\Theta = 90^\circ$ (bottom panels). 
    Left column: 1D horizontal profiles of $B'_z$ using different values of the $\sigma = 1$, $10$, and $20$ (green, blue, and red lines respectively) for $N_x=32$.
    Central and right columns: L1 norm errors of $B^* = \vec{B}\cdot\hvec{n}$ and $E^*=\vec{E}\cdot(\hvec{n}\times\hvec{e}_x)$ as functions of the resolution. 
    The dashed lines give the ideal convergence slope for the $4^{\rm th}$
    order.}
    \label{fig:te}
\end{figure*}

In the fluid rest frame, the resistive RMHD equations can be employed to study the propagation of light waves in the presence of a finite conductivity, as originally presented in \citet{Mignone_etal_2018} and paper I.
For a large plasma inertia ($\rho\to\infty$), only the Maxwell equations may be evolved in time:
\begin{equation}\label{eq:Maxwell_Telegraph}
  \left\{\begin{array}{l}
     \DS\pd{\vec{B}}{t} + \nabla\times\vec{E} = 0
     \\ \noalign{\medskip}
     \DS\pd{\vec{E}}{t} - \nabla\times\vec{B} = -\sigma\vec{E} \,.
  \end{array}\right.
\end{equation}
Eqns. (\ref{eq:Maxwell_Telegraph}) admit damped propagating plane wave solutions when $\sigma < 2|k|$, where $k$ is the wavenumber.
The exact mode in one dimension is 
\begin{equation}\label{eq:tg_exact}
  \begin{array}{l}
    \vec{B} =\DS B_1 e^{-\sigma t/2}\cos \phi(x,t) \hvec{n}
    \\ \noalign{\medskip}
    \vec{E} =\DS B_1 e^{-\sigma t/2}
               \left[  \frac{\mu}{k}     \cos\phi(x,t)
                      +\frac{\sigma}{2k} \sin\phi(x,t)\right]
                \hvec{n}\times\hvec{e}_x       \,,
    \end{array}   
\end{equation}
where $\phi(x,t) = kx - \mu t$, $\mu = (k^2 - \sigma^2/4)^{\HALF}$, while $B_1$ is the initial perturbation amplitude.
Here, in addition to paper I, we have introduced $\hvec{n} = (0,\, \cos\Theta,\, \sin\Theta)$ as the oscillation direction of the magnetic field in the $yz$ plane.

Following paper I, we rotate the 1D solution by an angle $\alpha$ around the $z$-axis and consider the $2D$ computational domain $x'\in [0,L_x]$, $y'\in [0,L_y]$ with $L_x=1$, $L_y=1/2$ discretized with $N_x\times N_x/2$ zones.
Solving in a 2D domain allows us to assess the correct discretization of multi-dimensional terms.
Electric and magnetic field vectors in this frame ($\vec{E}'$ and $\vec{B}'$) are obtained through the rotation matrix
\begin{equation}
  \{\vec{E},\vec{B}\}' = \left(\begin{array}{lll}
     \cos\alpha  & -\sin\alpha  &  0     \\ \noalign{\medskip}
     \sin\alpha  &  \cos\alpha  &  0     \\ \noalign{\medskip}
              0  &           0  &  1
  \end{array}\right)\{\vec{E},\vec{B}\} \,.
\end{equation}
The wavevector has orientation $\vec{k}= k_x(1,\tan\alpha,0)$, where $k_x=2\pi/L_x$ while $\tan\alpha = L_x/L_y = 2$.
Eq. (\ref{eq:tg_exact}) with $B_1=1$ and $\phi(\vec{x}',0)=\vec{k}\cdot\vec{x}'$ is used to initialize the electric and magnetic field.
We also set density to a very large value ($\rho=10^{12}$) while velocity are set to zero.
Periodic boundary conditions are imposed everywhere.

We employ the $4^{\rm th}$-order ARK4 IMEX scheme and run computations at different grid resolutions for exactly one wave period $T=2\pi/\mu$ where $\mu$ is defined after Eq. (\ref{eq:tg_exact}).
In Fig. \ref{fig:te} we plot, from left to right, the profiles of $B'_z$ (at the resolution $N_x=32$), the $L_1$-norm errors of $\vec{B}\cdot\hvec{n}$ and $\vec{E}\cdot(\hvec{n}\times\hvec{e}_x)$ as a function of the grid resolution (vectors are transformed back to the 1D frame), for $\Theta = 30^\circ$ (top panels) and $\Theta = 90^\circ$ (bottom panels).
Green, blue and red colors refer to computations obtained with $\sigma = 1, 10$ and $20$, respectively.
In all cases we recover the expected order of accuracy (4).

\subsection{Stationary Charged Vortex}

\begin{figure*}
  \centering
  \includegraphics[width=0.7\textwidth]{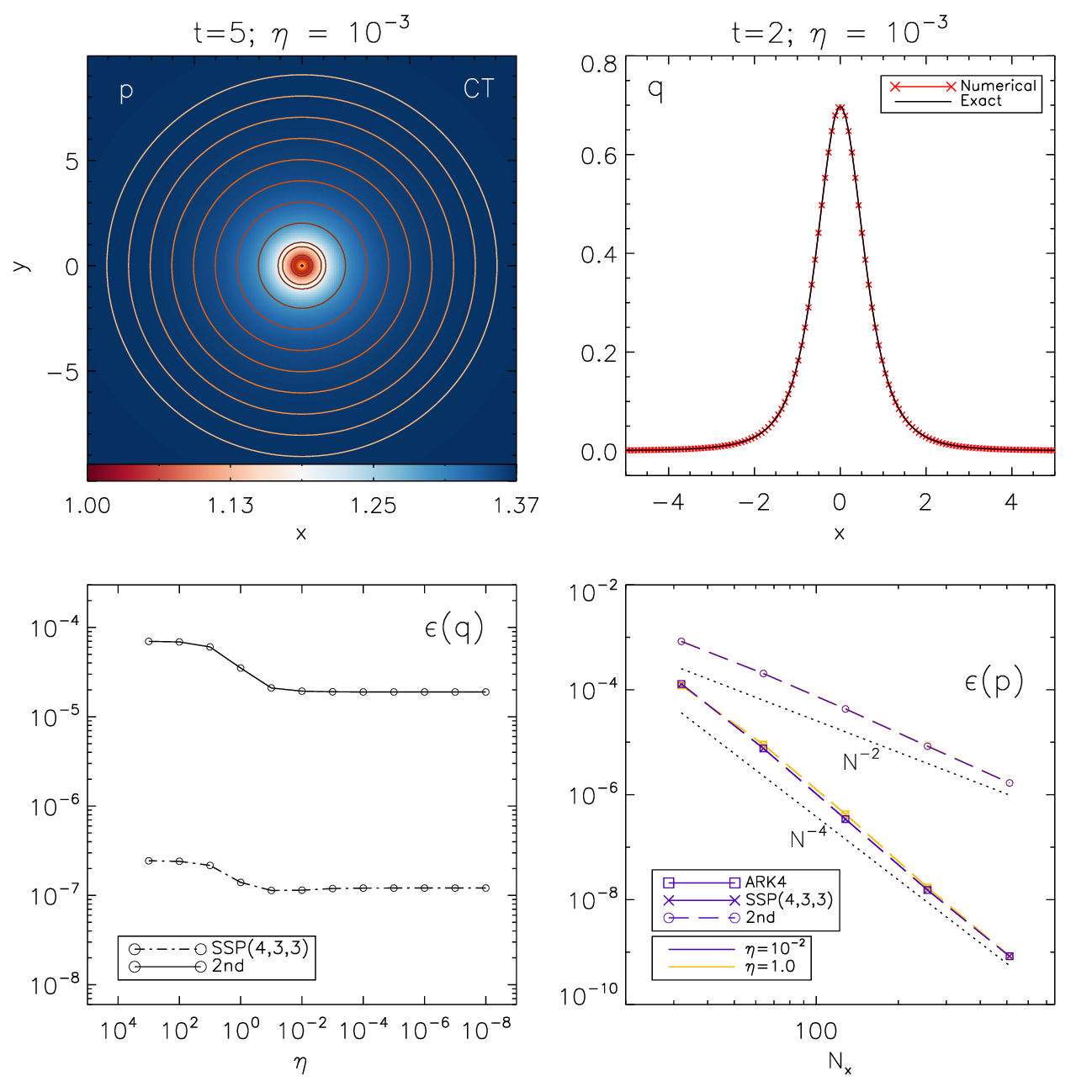}
  \caption{ \small Numerical results of the charged vortex problem. Top-left panel: we show 8 equally spaced contour levels of $E$ chosen as $(q_0 /2)r_k /(r_k^2 + 1)$ (where $r_k = 1,...,8$) overlaid on the coloured maps of pressure at $t = 5$. 
  Resistivity is set to $\eta = 10^{-3}$ and the grid resolution is $256^2$ zones.
  Top-right panel: 1D horizontal cuts, at $y = 0$, of charge at $t = 3$ obtained with CT. 
  Bottom-left panel: $L_1$ norm errors of the charge as functions of the resistivity using CT with $256^2$ zones. 
  Bottom-right panel: $L_1$ norm errors of the gas pressure as functions of the resolution for the SSP(4,3,3), ARK4 and the second-order scheme.
  The dotted lines gives the expected second order scaling.
  }
  \label{fig:charged_vortex}
\end{figure*}

We now consider an exact equilibrium solution of the full resistive RMHD equations, originally introduced in paper I.
It consists of a rotating charged flow embedded in a vertical magnetic field.
Assuming purely radial dependence, hydrodynamic and electromagnetic variables are best expressed in a cylindrical coordinate system ($r,\,\phi,\,z$) as
\begin{equation}\label{eq:charged_vortex}
  \left\{\begin{array}{l}
   E_r      = \DS \frac{q_0}{2}\frac{r}{r^2+1}
   \\ \noalign{\medskip}
   B_z      = \DS \frac{\sqrt{(r^2 + 1)^2 - q_0^2/4}}
                           {r^2 + 1} 
   \\ \noalign{\medskip}
   v_{\phi} = \DS -\frac{q_0}{2}\frac{r}{\sqrt{(r^2+1)^2 - q_0^2/4}}
   \\ \noalign{\medskip}
   p        = -\DS \frac{\rho}{\Gamma_1} + \left[
                   \frac{4r^2 + 4 - q_0^2}
                        {(r^2 + 1)(4 - q_0^2)}\right]^{\Gamma_1/2}
                   \left(p_0 + \DS\frac{\rho}{\Gamma_1}\right)
   \\ \noalign{\medskip}
   q        = \DS\frac{q_0}{(r^2+1)^2} \,,
  \end{array}\right.
\end{equation}
where $\Gamma_1 = \Gamma/(\Gamma-1)$, $p_0=0.1$, $q_0=0.7$ is the charged at $r=0$.
Density is set to unity. 
The initial equilibrium condition given by Eq. (\ref{eq:charged_vortex}) is mapped to a 2D Cartesian domain ($x,y\in[-10,10]$) and the system is evolved until $t=5$.
Boundary conditions replicate the equilibrium solution throughout the integration. 
Since the equilibrium condition does not depend on the resistivity, numerical computations with different values of the resistivity $\eta$ depend solely on the stability of the algorithm.

For the present purpose we have tested both the $3^{\rm rd}$-order SSP3(4,3,3) RK-IMEX as well as the $4^{\rm th}$-order ARK4 schemes.
Results are shown in the four panels of Fig. \ref{fig:charged_vortex}.
The integrity of the solution is effectively maintained. 
This is demonstrated by both the pressure color map (top left panel) showing also iso-contour levels of the electric field magnitude) and the 1D horizontal cuts of the charge (top right).

In the bottom left panel of Fig. \ref{fig:charged_vortex} we plot the charge error as a function of the resistivity and compare our results with those of paper I using a $2^{\rm nd}$-order method.
Our $4^{\rm th}$-order scheme is able to reduce the error  by two orders of magnitude and we have found that both the SSP(4,3,3) and the ARK4 yield nearly identical results. 
However, the ARK4 method becomes unstable for $\eta \lesssim 10^{-4}$.
For this reason, we plot only the results obtained with the SSP method.
Overall, the charge error exhibits minimal variation over time, underscoring the stability of the algorithm for any value of the resistivity parameter in the chosen range. 

In the lower right quadrant of Fig. \ref{fig:charged_vortex}, we present the L1-norm errors of gas pressure against resolution, showcasing the achievement of $4^{\rm th}$-order accuracy for both methods.
Notably, in comparison with the results presented in Paper I, both charge and pressure errors are markedly reduced, with a two-order improvement in convergence.

\subsection{2D Relativistic Rotor}

\begin{figure*}
  \centering
  \includegraphics[width=0.8\textwidth]{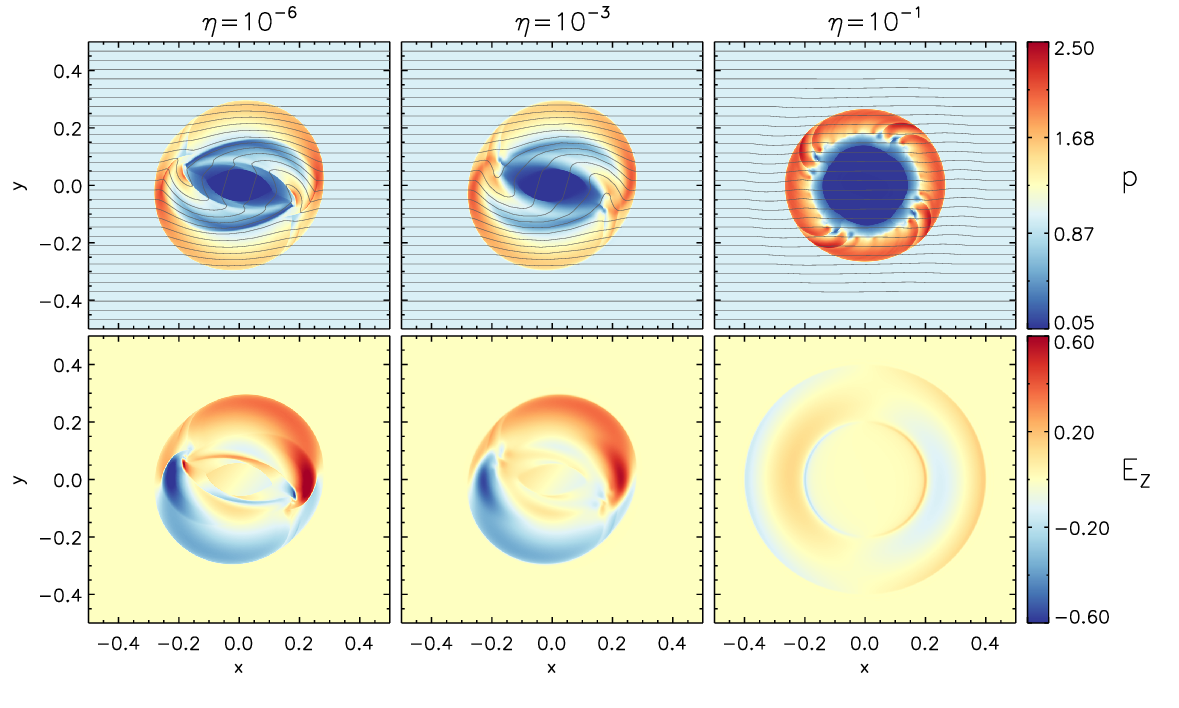}
  \caption{\small Snapshots of the gas pressure $p$ (upper panels), and profiles of $E_z$ (lower panels) for the resistive rotor test at its final time $t = 0.3$ at $400 \times 400$ grid points using the SSP3(4,3,3) IMEX time-stepping.
  From left to right, results correspond to computations with different resistivities $\eta = 10^{-6}$, $\eta = 10^{-3}$ and $\eta = 10^{-1}$.
  }
  \label{fig:rotor}
\end{figure*}

We propose a resistive variant of the ideal special relativistic MHD test proposed by \citet{DelZanna_etal2003}, as also presented by \citet{Dumbser_Zanotti2009, Bucciantini_delZanna2013, Miranda_etal2018, Nakamura_2023}.
An overdense clump ($\rho = 10$) concentrated in a circular region ($r = 0.1$) at the center of the computational square $[-0.5, 0.5]^2$ is rapidly spinning ($\omega = 8.5$) in a uniform medium ($\rho = 1$, $p = 1$, $\Gamma = 4/3$) at rest.
The vortex is embedded in a constant magnetic field $\Vec{B} = (1, 0, 0)$ which is progressively wrapping around the spinning vortex, launching torsional Alfv\'en waves in the ambient fluid.
At the final time ($t = 0.3$), the torsional Alfv\'en waves have approximately been rotated of a $90 \degree$ angle, with the magnetic pressure deforming the initially circular vortex into its characteristic oval shape.

We perform a set of $400 \times 400$ grid points simulations with increasing resistivity, from a quasi-ideal regime ($\eta = 10^{-6}$, $10^{-3}$) to a fully resistive one ($\eta = 10^{-1}$).
Both the SSP(4,3,3) and the ARK4 were used for this test using order reduction to WENO3.
As the resistivity increases, the vortex angular momentum as well as the torsional Alfv\'en waves that cause the well known deformation are dissipated, leading to a more circular shape.
Fig. \ref{fig:rotor} presents the snapshots of thermal pressure $p$ (top panels), and the electric field $E_z$ (bottom panels) at the final time $t=0.3$.
While results for SSP(4,3,3) appear essentially identical to those obtained with ARK4, only the former is able to succesfully complete the integration with $\eta = 10^{-6}$, again confirming the reduced stability of the $4^{\rm th}$-order method, which is not strong stability preserving.

It is relevant to notice how the pressure distribution is affected by the onset of spurious modes ($m=4$) in the most resistive regime ($\eta = 10^{-1}$).
This artifact results from the employment of Cartesian coordinates and it becomes more evident when using a MHLLC solver rather than a more diffusive one (with which the effect is substantially mitigated).

\subsection{3D Blast Wave}

\begin{figure*}
  \centering
  \includegraphics[width=0.9\textwidth]{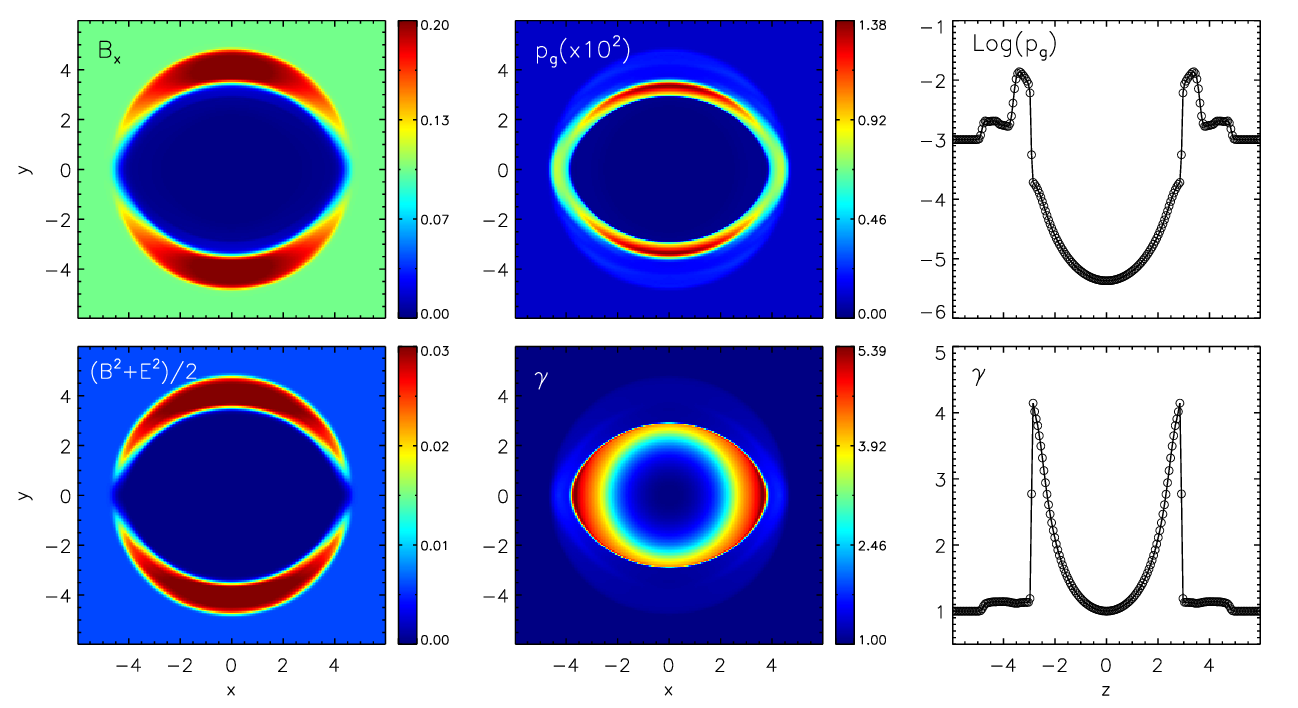}
  \caption{\small 3D Blast wave at $t = 4$ using $\eta = 10^{-6}$ and the CT scheme. 
  In the top panels we show 2D slices in the $xy$ plane of $B_x$ (left), gas pressure $p_g$ (middle) and its 1D profile along the z-axis (right).
  In the bottom panels we show 2D slices of the electromagnetic pressure (left), Lorentz factor (middle) together with its cuts of along the z-axis (right).
  }
  \label{fig:blast3D}
\end{figure*}

The magnetized blast wave problem is used here to assess the robustness of the numerical method in handling the propagation of shock wave through magnetic field of varying strength as well as the capacity of preserving the symmetric properties of the solution. 
In fact, unphysical features could emerge if the divergence-free condition is not properly controlled \citep[see, e.g.][]{Mignone_Bodo2006, DelZanna_etal2007}.

Following paper I, we consider the computational domain $x,y,z\in[-6,6]$ threaded by a uniform magnetic field $\vec{B}= B_0\hvec{e}_x$ with $B_0=0.1$.
The plasma is initially at rest while density and pressure experience a sharp transition across a sphere of radius $r_c=0.8$
\begin{equation}
  \left(\begin{array}{l}
      \rho \\ \noalign{\smallskip}
      p    
  \end{array}\right)
     = 
  f_t(r)\left(\begin{array}{l}
      \rho_{\rm in} \\ \noalign{\smallskip}
      p_{\rm in}    
  \end{array}\right)
     + \left(1 - f_t(r)\right)
  \left(\begin{array}{l}
      \rho_{\rm out} \\ \noalign{\smallskip}
      p_{\rm out}
  \end{array}\right)
\end{equation}
where $f_t(r)$ is a taper (smoothing) function, 
\begin{equation}
  f_t(r) = \max\left[\min\left(\frac{e^\chi - e}{1 - e},1\right),0\right] \,,
\end{equation}
with $\chi = (r-r_c)/(1-r_c)$.
Similar 3D configurations have been used by \cite{Komissarov_2007} and \cite{Dionys_etal2013} although with larger resistivity or lower magnetic fields. 
The system is evolved until $t=4$ using the SSP(4,3,3) method. 
We also enable order reduction to linear for this test.
Zero-gradient boundary conditions apply everywhere.

Fig. \ref{fig:blast3D} shows the outcomes of the simulation, presenting, in the top panel, 2D maps of the $x$-component of magnetic field (left), gas pressure (middle) while, in the bottom panel, electromagnetic pressure (left) and Lorentz factor (middle). 
In the rightmost column, we show one-dimensional profiles along the $z$-axis of thermal pressure (top) and Lorentz factor (bottom).
As a consequence of the explosion, a swiftly advancing forward shock moves in an (almost) radial manner, while concurrently, a trailing reverse shock marks the boundary of the internal region where radial expansion occurs. 
Electromagnetic fields pile up in the $y$-direction, forming a shell characterized by elevated magnetic pressure. 
Gas exhibits a pronounced preference for motion along the $x$-direction, attaining a heightened Lorentz factor ($\gamma_{\max} \approx 5.4$).

\subsection{Tearing mode}

In what follows we validate our method on the linear growth of the ideal tearing mode instability with a Harris current sheet profile, following the results presented by   \cite{DelZanna_etal_2016} \citep[see also][and paper I]{Miranda_etal2018}.

In this problem, a static ($v = 0$) plasma is uniformly distributed in the computational domain with constant initial density and pressure $\rho_0$ and $p_0$.
The initial magnetic field profile satisfies the force-free condition and it is given by
\begin{equation}
    \textbf{B} = B_0 \left[ \tanh \left( \frac{x}{a} \right) \hvec{e}_y + \sech \left( \frac{x}{a} \right) \hvec{e}_z \right] \, ,
\end{equation}
while the initial electric field is null everywhere.
The physical parameters that set the proper conditions for the instability to grow are the current sheet thickness $a$, 
the magnetization $\Sigma = B_0^2 / \rho_0$, the plasma beta $\beta = 2p_0 /B_0^2$, the Alfvén velocity $v_a = B_0 /\sqrt{B_0^2 + w_0} = B_0 /\sqrt{B_0^2 + \rho_0 + 4p_0}$ obtained by assuming an ideal gas law with relativistic adiabatic index $\Gamma = 4/3$, and the Lundquist number $S = v_a L/\eta$, where $\eta$ is the resistivity and $L$ is the reference spatial scale of variation.
In order to trigger the instability, $S\gg 1$.

\begin{figure*}
  \centering
  \includegraphics[width=1\textwidth]{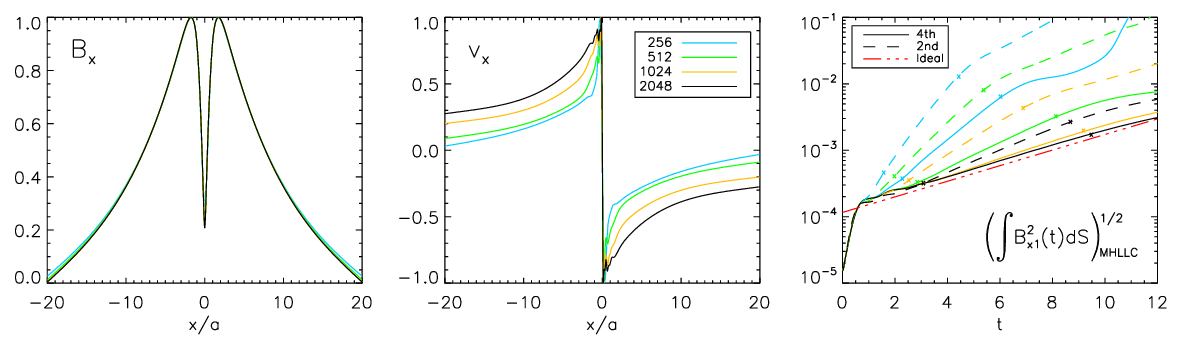}
  \caption{\small Horizontal cuts of the x- component of the magnetic field (left panel), and of the velocity (middle panel) for the tearing mode instability at $t = 10$.  
  The profiles are normalized to unity and show the trends along $y = 0$ for the magnetic field (left panel), and along $\frac{3}{4}L_y$ for the velocity (middle panel). 
  Each resolution ($N_x = \{256, 512, 1024, 2048\}$) is demarked with a different color.
  Measured growth rates (right panel) are plotted as solid (dashed) lines for $4^{\rm th}$- ($2^{\rm nd}$-) order data and are compared to the ideal trend (dashed dotted line).
  Colorful crosses mark the time intervals for which each growth rate has been calculated.
  }
  \label{fig:tearing}
\end{figure*}

Generalizing the MHD results obtained by \cite{Pucci_Velli_2014} and \cite{Landi_etal2015} to this regime, provided a sufficiently thin current sheet, the reconnection process is expected to occur on the ideal Alfv\'enic time scale $\tau_a = L/v_a$.
The critical (inverse) aspect ratio determining this regime is
\begin{equation}
    a = S^{-1/3} L \, .
\end{equation}
In the limit of large $S$, the growth rate $\gamma_{\rm TM}$ of the instability becomes $\gamma_{\rm TM} \simeq 0.6 v_a/L$, which, in a relativistic regime ($v_a \rightarrow c = 1$), results in a very efficient process commonly referred to as \textit{ideal} tearing instability.

The initial equilibrium is perturbed by a small amplitude ($\epsilon = 10^{-4}$) variation in the magnetic field employing only one wavenumber $k$
\begin{equation}
\DS \delta \textbf{B} = \epsilon B_0 \sech \left(\frac{x}{a} \right)
  \left(\begin{array}{c}
   \DS  \cos(ky) \\ \noalign{\smallskip}
   \DS  \frac{\sin(ky)}{ka} \tanh\left(\frac{x}{a} \right) \\ \noalign{\smallskip}
   \DS  0    
  \end{array}\right) \, .
\end{equation}
Following \S 5.5 of \cite{Mignone_etal_2019}, we initialize the magnetic field in the $xy$ plane with the vector potential
\begin{equation}
    \DS A_z = -B_0 \left[ a \log\left(\cosh \frac{x}{a} \right) - \frac{\epsilon}{k} \sin(ky)\sech\left(\frac{x}{a} \right) \right] \, .
\end{equation}

In accordance to the work developed by \cite{DelZanna_etal_2016} and later resumed by \cite{Mignone_etal_2019}, the linear theory predicts that the wavenumber associated with the fastest growing mode of the ideal tearing instability is $k_{\rm max} = 1.4 S^{1/6} = 14$.
The corresponding growth rate is $\gamma_{\rm TM} \simeq 0.3$.
Nonetheless, due to the neglection of the compressible terms in the linear approximation as presented in \cite{DelZanna_etal_2016}, the maximum value of the dispersion relation is shifted to $k = 12$ with the consequent modification of the expected growth rate to $\gamma_{\rm TM} \simeq 0.27$.
We therefore align with the works of \cite{DelZanna_etal_2016}, \cite{Miranda_etal2018}, and \cite{Mignone_etal_2019} and choose $k = 12$ for our simulations.
Likewise, we set $L = B_0 = \Sigma = \beta = 1$, deriving therefore $\rho_0 = 1$, $p_0 = 0.5$, $v_a = 0.5$, and $S = 10^6$.
In this way $a = 0.01$, and $ \eta = 5 \times 10^{-7}$. 
In order to recover the predicted growth rate, we perform a set of simulations with increasing resolution $N_x \times N_x/4$, ($N_x = \{256, 512, 1024, 2048\}$) on a rectangular computational domain with $x \in [-20a, 20a] = [-0.2, 0.2]$ and outflow boundary conditions, and $y \in [0, 0.523598]$ imposing periodicity along $y$.
Simulations have been carried out until $t = 20$ using the IMEX SSP(4,3,3) time-stepping.

In the left and middle panels of Fig. \ref{fig:tearing} we  plot the horizontal cuts at $y=0$ of the $x$-components of the magnetic filed (left panel), velocity (middle panel) at $t=10$ for all resolutions.
The magnetic field at all resolutions closely resembles the eigenfunction solutions presented by \cite{DelZanna_etal_2016} (see Fig. 1 of that paper).
The same holds for the velocity profile, although a more pronounced dependency on resolution is observed along with some clipping.
Following \cite{Miranda_etal2018}, and \cite{Mignone_etal_2019} we measure the growth rate by first computing 
\begin{equation} \label{eq:growth}
    \gamma(t) = \DS \frac{1}{2}\log\left(\int B^2_x(t) dS \right) \,,
\end{equation}
which we plot, as a function of time, in the right panel of Fig. \ref{fig:tearing} (solid lines) along with $2^{\rm nd}$-order data from \cite{Mignone_etal_2019} (dotted lines) for each resolution.
The growth rate is then computed from a linear fit of Eq. (\ref{eq:growth}) while the red dashed-dotted line represents the ideal value ($\gamma_{\rm TM} \simeq 0.27$).
Table \ref{tab:growth_rates} compares the results obtained with the $2^{\rm nd}$- and $4^{\rm th}$-order scheme (middle and right columns, respectively).
For each slope, a different time-window has been chosen in order to properly fit the linear growth of the fastest growing mode (crosses in Fig. \ref{fig:tearing}) during the first linear phase.
While both the $2^{\rm nd}$- and $4^{\rm th}$-order schemes show systematic convergence towards the predicted value, the latter requires approximately half of the resolution (or less) to reproduce the results of the former.
The best agreement is reached with the $4^{\rm th}$-order scheme using $2048$ zones.

\begin{table}
\centering
\begin{tabular}{ |p{1.5cm}||p{2cm}||p{2cm}|}
 \hline
 Resolution & $2^{\rm nd}$-order & $4^{\rm th}$-order \\
 \hline
 \hline
$256\times64$     &  1.19  &  0.75  \\
$512\times128$    &  0.89  &  0.44  \\      
$1024\times256$   &  0.58  &  0.30  \\
$2048\times512$   &  0.38  &  0.27  \\
 \hline
\end{tabular}
   \caption{Growth rates of the linear phase of the tearing instability as measured from $2^{\rm nd}$-order and $4^{\rm th}$-order simulation data, listed by resolution. 
   The measured growth rates refer to the onset of the first linear phase of the instability.
   The time intervals chosen to evaluate each slope are marked as crosses in the right panel of Fig. \ref{fig:tearing}.}
    \label{tab:growth_rates}
\end{table}

\subsection{3D Kelvin-Helmholtz Instability}

\begin{figure*}
    \centering
    \includegraphics[width=0.9\textwidth]{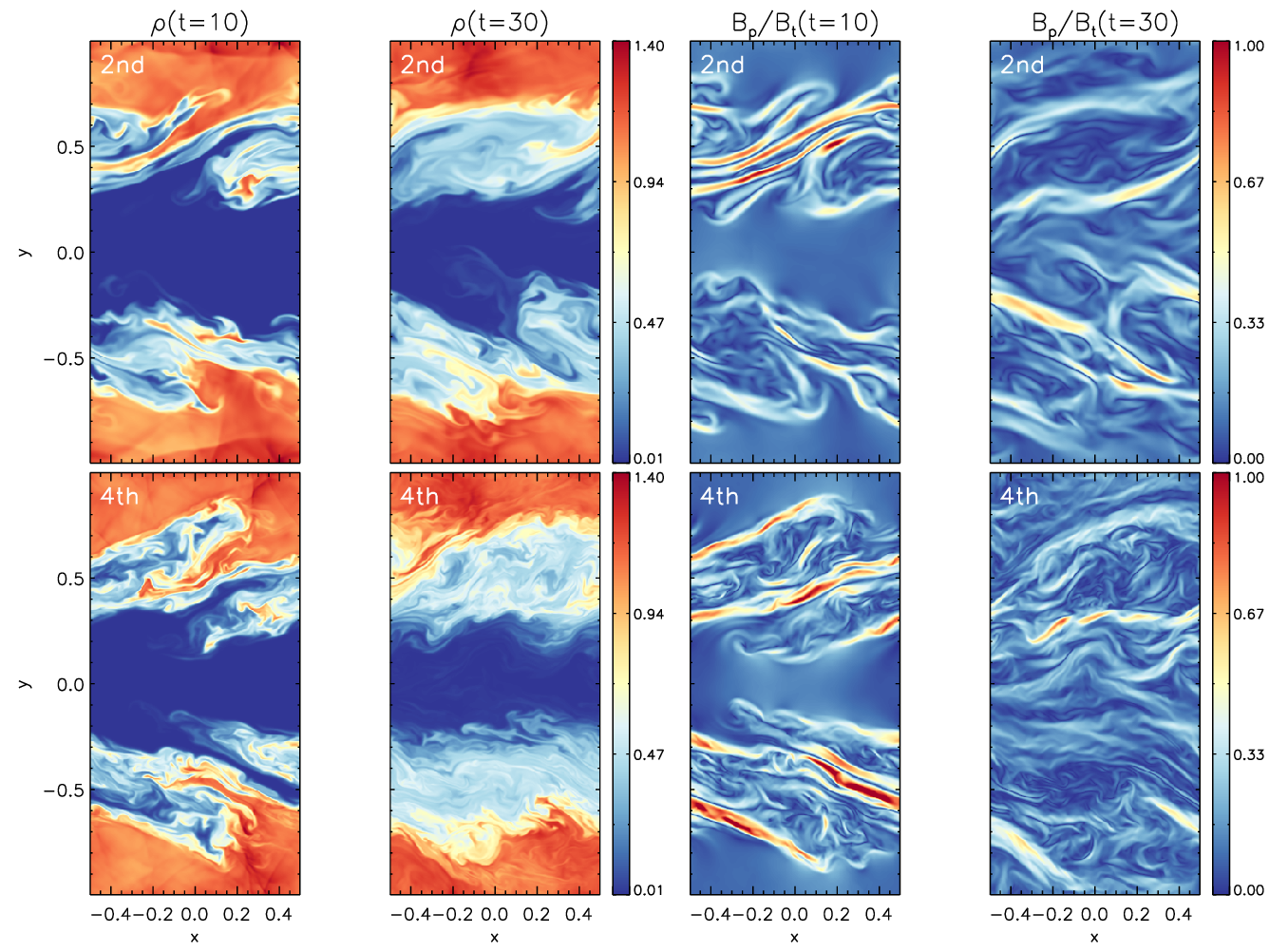}
    \caption{\small 2D slice-cut of the density and poloidal to toroidal magnetic field at $z=0$ for the 3D Kelvin-Helmholtz instability at $t = 10$ and $30$ at the resolution of $256\times512\times256$ zones.
    Here $B_p = \sqrt{B_x^2 + B_y^2}$.}
    \label{fig:kh3D_evolution}
\end{figure*}

\begin{figure*}
    \centering
    \includegraphics[width=0.9\textwidth]{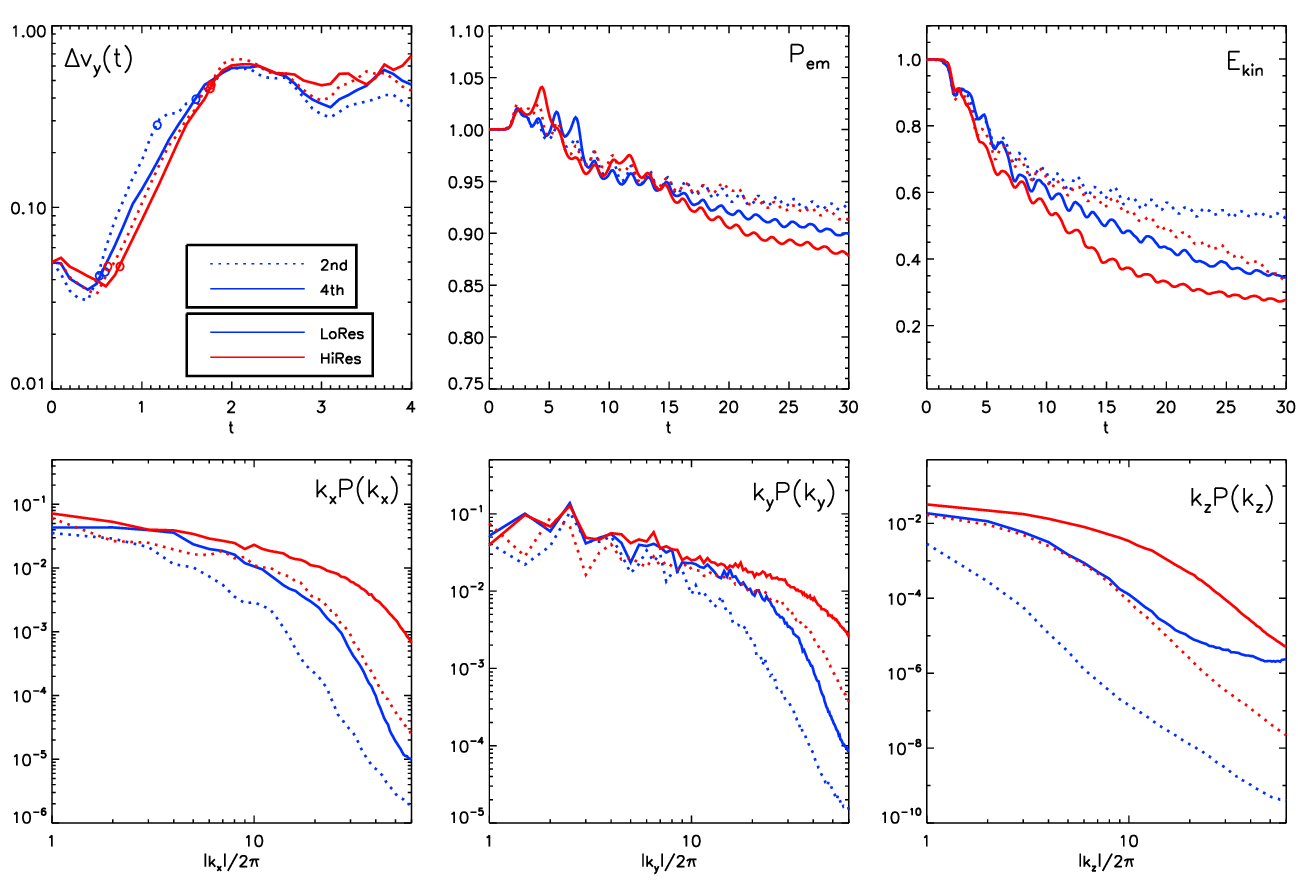}
    \caption{\small Top panels: growth rate (left) and volume-integrated electromagnetic $P_{\rm em} = (\vec{B}^2+\vec{E}^2)/2$ (middle) and kinetic $E_{\rm kin} =  \rho\gamma(\gamma-1)$ (right) as a function of time for the 3D Kelvin-Helmholtz instability.
    Circles in the top-left plot mark the time windows used to compute the growth rate.
    Bottom panels: from left to right, spectral energy densities in the $x$, $y$ and $z$-direction at $t = 15$.
    Solid and dashed lines are used for the $4^{\rm th}$ and $2^{\rm nd}$ order schemes while blue and red are used for the resolution of $128\times 256\times128$ and $256\times 512\times256$ zones, respectively. }
    \label{fig:kh3D_growth}
\end{figure*}

As a final application benchmark, we consider a 3D version of Kelvin-Helmholtz instability {\bf in relativistic MHD \citep[see, e.g.][]{Bucciantini_delZanna2006} } as presented in paper I and earlier, in 2D, also by \cite{Mizuno2013, Beckwith_Stone2011}.
At $t=0$, we prescribe a double shear layer configuration with non-uniform density and constant magnetic field:
\begin{equation}
  \begin{array}{l}
  \DS \rho = \frac{1+\tanh\phi(y)}{2}\rho_h + \frac{1-\tanh\phi(y)}{2}\rho_h  \,,
  \\ \noalign{\medskip}
  \DS \vec{v} =   v_{\rm sh}\tanh\phi(y)\hvec{e}_x
                + A_0\left[{\rm sign}(y)\sin(2\pi x)\hvec{e}_y 
                            + \zeta \hvec{e}_z\right] f(y)  \,,
  \\ \noalign{\medskip}
  \DS \vec{B} = \left(\sqrt{2p\sigma\mu_p},\,0,\, \sqrt{2p\sigma\mu_t} \right)  \,,\\ \noalign{\medskip}  
  \end{array}
\end{equation}
where $v_{\rm sh} = 1/2$ is the shear velocity, $a = 0.01$ is the shear thickness, $\phi(y) = (|y|-1/2)/a$ while $A_0=0.1$ is the perturbation amplitude and $\zeta$ is a random number in the range $[-1,1]$.
The function 
\begin{equation}
  f(y) = e^{-\phi^2a^2/\alpha^2}
\end{equation}
(with $\alpha=0.1$) is used to damp the perturbations as we move away from the two interface located at $y=\pm 1/2$.
The parameters $\mu_p$ and $\mu_t$ control the in plane  ($\mu_p = 0.01$) and out of plane ($\mu_t = 1$) magnetic field strength.
We set resistivity to $\eta = 1/\sigma=10^{-5}$ and evolve the system until $t=30$ using the MHLLC Riemann solver.
Following paper I, we perform simulations on the Cartesian domain $x,z\in[-1/2,1/2]$ and $y\in [-1,1]$ with uniform grid resolution $N\times 2N\times N$, with $N=128$ (low-resolution) and $N=256$ (high resolution).

Simulations have been repeated with both the high-order scheme - employing IMEX SSP3(4,3,3) time-stepping - as well as using also the $2^{\rm nd}$-order scheme presented in paper I, based on piecewise linear reconstruction and second-order IMEX time-stepping.
For the $4^{\rm th}$-order method we employ order reduction (linear).

The time evolution of the instability can be observed in Fig. \ref{fig:kh3D_evolution} where we show 2D slices at $z=0$ for density ($1^{\rm st}$ and $2^{\rm nd}$ raw from top) and poloidal to toroidal magnetic field ($3^{\rm rd}$ and $4^{\rm th}$) with $N=256$.
The large-scale motion remains confined along the initial shear direction and the turbulent state is characterized by sheet-like thin structures, where most of the magnetic field energy is trapped. 
These rounded slabs remain roughly parallel to the $xz$ plane although significant medium scale structures develop in this direction for $t\gtrsim 20$ around the shear layer.

A direct measure of the growth rate, quantified by $\Delta v_y = (\max(v_y) - \min(v_y))/2$, is shown in the top-left panel of Fig.  \ref{fig:kh3D_growth} for both the $2^{\rm nd}$ and $4^{\rm th}$-order simulations at low and high resolutions (blue and red colors, respectively).
As noticed in paper I (see Fig. 11 of that paper), with the MHLLC Riemann solver, an increase in resolution (or scheme accuracy) is accompanied by a gradual reduction of the growth rate, until convergence is finally reached.
This is also the case here and clearly visible with the $2^{\rm nd}$-order method (red lines) but only marginally with the $4^{\rm th}$-order scheme which shows better convergence properties at low resolution. 
A quantitative measure of the growth rate, in the range given by the circle symbols in the figure, yields indeed $\approx 2.94$ and $\approx 2.14$ for the $2^{\rm nd}$-order scheme (low and high resolution, respectively) and $\approx 2.17$ and $\approx 2.31$ for the $4^{\rm th}$-order scheme.

The enhanced dissipation of electromagnetic and kinetic energies (top middle and right panels in Fig. \ref{fig:kh3D_growth}) using our high-order scheme stems from the increased turbulence that comes from having an effective dissipation at scales smaller than the low-order case.
In this sense, the performance of our high-order scheme at low resolutions ($N_x=128$) already outperforms the one obtained by the $2^{\rm nd}$-order scheme at twice the resolution.

The amount of small-scale features is an indirect measurement of the numerical dissipation, which we quantify through the spectral energy density,
\begin{equation}\label{eq:KH_fourier}
    P(k_x) = \int |{\cal F}(k_x,y,z)|^2 \, dy\,dz\,,
\end{equation}
where ${\cal F}(k_x,y,z)$ is the 1D fast Fourier transform of the density taken across the $x$ dimension.
$P(k_y)$ and $P(k_z)$ are constructed similarly, by corresponding index permutations.
Results, shown in the bottom rows of Fig. \ref{fig:kh3D_growth} ($P(k_y)$ and $P(k_z)$), indicate that large-scale modes are better resolved with the $4^{\rm th}$-order scheme.

\section{Summary}
\label{sec:summary}
%
%

We have presented a genuinely $4^{\rm th}$-order finite volume algorithm for the solution of the resistive relativistic MHD equations in multiple dimensions. 
To our knowledge, this is one of the few contributions addressing the design of $4^{th}$-order methods based on an IMEX approach for multidimensional PDEs with stiff terms.

High-order quadrature rules to obtain volume, surface or line averages are evaluated using appropriate Simpson rules, as outlined in the seminal paper by \cite{Corquodale_Colella2011}.
Point values are recovered from volume (or surface) averages following the opposite procedure.
Fourth-order spatial accuracy is achieved by reconstructing point values (rather than 1D volume averages) using the modified WENO-Z reconstruction of \cite{Berta_etal_2024}, while for temporal integration, we have implemented and compared the $3^{\rm rd}$-order Runge Kutta IMEX SSP3(4,3,3) scheme of \cite{Pareschi_Russo2003, Pareschi_Russo2005} with the $4^{\rm th}$-order IMEX-ARK4(3)7L[2]SA\_1 scheme of \cite{Kennedy_Carpenter2019} (ARK). 
The divergence-free condition is ensured by evolving magnetic fields according to the constrained transport formalism, while electric field is treated as a volume average quantity.
Equations are evolved in conservation form, requiring the solution of a Riemann problem between discontinuous left and right states at zone interfaces.
Our scheme employs only 3 ghost zones, and thus a more compact stencil when compared to its previous version \citep{Berta_etal_2024}.

Numerical benchmarks in two and three dimensions demonstrate the accuracy of the method along with reduced numerical dissipation.
For smooth problems our results show superior accuracy and faster convergence when compared to standard $2^{\rm nd}$-order schemes.  
While the ARK4 retains the temporal $4^{\rm th}$-order accuracy for smooth problems, we found in our experience that the $3^{\rm rd}$-order SSP scheme is more robust for lower values of the resistivity, $\eta \lesssim 10^{-5}$ and thus preferable for highly stiff problems.

The scheme can also cope with complex flow features including steep gradients and discontinuous waves, although the spatial order of the scheme may have to be locally reduced (\quotes{fallback} approach) so as to enhance robustness in critical regions. 
To identify such regions, we employ the discontinuity detector proposed in \cite{Berta_etal_2024}, based on the ratio of derivative, specifically designed to detect high-frequency oscillations typically arising in proximity of a discontinuous front.

The employment of genuinely $4^{\rm th}$-order methods may represent a major upgrade in the field of astrophysical plasma computations, allowing more accurate computations at a reduced cost.

\section*{Acknowledgements}

\noindent
Computational resources were provided by the Centro di Competenza sul Calcolo Scientifico (C3S) of the University of Torino (\url{c3s.unito.it}).

\vspace{7pt}\noindent
This project has received funding from the European Union's Horizon Europe research and innovation programme under the Marie Sk\l{}odowska-Curie grant agreement No. 101064953 (GR-PLUTO).

\vspace{7pt}\noindent
This work has received funding from the European High Performance Computing Joint Undertaking (JU) and Belgium, Czech Republic, France, Germany, Greece, Italy, Norway, and Spain under grant agreement No 101093441 (SPACE).

\vspace{7pt}\noindent
This paper is supported by the  Fondazione ICSC, Spoke 3 Astrophysics and Cosmos Observations. 
National Recovery and Resilience Plan (Piano Nazionale di Ripresa e Resilienza, PNRR) Project ID CN\_00000013 \quotes{Italian Research Center on High-Performance Computing, Big Data and Quantum Computing}  funded by MUR Missione 4 Componente 2 Investimento 1.4: Potenziamento strutture di ricerca e creazione di \quotes{campioni nazionali di $R\&S$ (M4C2-19 )} - Next Generation EU (NGEU).

\vspace{7pt}\noindent
The partial support from Gruppo Nazionale Calcolo Scientifico-Istituto Nazionale di Alta Matematica (GNCS-INdAM) and MIUR-PRIN Project 2022, No. 2022KKJP4X \quotes{Advanced numerical methods for time dependent parametric partial differential equations with applications} is also acknowledged.

\section*{Data Availability}

The data underlying this article will be shared on reasonable request to the corresponding author.



\bibliographystyle{mymnras}
\bibliography{paper} 

\begin{thebibliography}{}
\makeatletter
\relax
\def\mn@urlcharsother{\let\do\@makeother \do\$\do\&\do\#\do\^\do\_\do\%\do\~}
\def\mn@doi{\begingroup\mn@urlcharsother \@ifnextchar [ {\mn@doi@}
  {\mn@doi@[]}}
\def\mn@doi@[#1]#2{\def\@tempa{#1}\ifx\@tempa\@empty \href
  {http://dx.doi.org/#2} {doi:#2}\else \href {http://dx.doi.org/#2} {#1}\fi
  \endgroup}
\def\mn@eprint#1#2{\mn@eprint@#1:#2::\@nil}
\def\mn@eprint@arXiv#1{\href {http://arxiv.org/abs/#1} {{\tt arXiv:#1}}}
\def\mn@eprint@dblp#1{\href {http://dblp.uni-trier.de/rec/bibtex/#1.xml}
  {dblp:#1}}
\def\mn@eprint@#1:#2:#3:#4\@nil{\def\@tempa {#1}\def\@tempb {#2}\def\@tempc
  {#3}\ifx \@tempc \@empty \let \@tempc \@tempb \let \@tempb \@tempa \fi \ifx
  \@tempb \@empty \def\@tempb {arXiv}\fi \@ifundefined
  {mn@eprint@\@tempb}{\@tempb:\@tempc}{\expandafter \expandafter \csname
  mn@eprint@\@tempb\endcsname \expandafter{\@tempc}}}

\bibitem[\protect\citeauthoryear{Ascher, Ruuth  \& Spiteri}{Ascher
  et~al.}{1997}]{Ascher1997}
Ascher U.~M.,  Ruuth S.~J.,   Spiteri R.~J.,  1997, \mn@doi [Applied Numerical
  Mathematics] {https://doi.org/10.1016/S0168-9274(97)00056-1}, 25, 151

\bibitem[\protect\citeauthoryear{{Balsara} \& {Spicer}}{{Balsara} \&
  {Spicer}}{1999}]{Balsara_Spicer1999}
{Balsara} D.~S.,  {Spicer} D.~S.,  1999, \mn@doi [Journal of Computational
  Physics] {10.1006/jcph.1998.6153}, \href
  {http://adsabs.harvard.edu/abs/1999JCoPh.149..270B} {149, 270}

\bibitem[\protect\citeauthoryear{{Beckwith} \& {Stone}}{{Beckwith} \&
  {Stone}}{2011}]{Beckwith_Stone2011}
{Beckwith} K.,  {Stone} J.~M.,  2011, \mn@doi [\apjs]
  {10.1088/0067-0049/193/1/6}, \href
  {https://ui.adsabs.harvard.edu/abs/2011ApJS..193....6B} {193, 6}

\bibitem[\protect\citeauthoryear{Berta, Mignone, Bugli  \& Mattia}{Berta
  et~al.}{2024}]{Berta_etal_2024}
Berta V.,  Mignone A.,  Bugli M.,   Mattia G.,  2024, \mn@doi [Journal of
  Computational Physics] {https://doi.org/10.1016/j.jcp.2023.112701}, 499,
  112701

\bibitem[\protect\citeauthoryear{{Borges}, {Carmona}, {Costa}  \&
  {Don}}{{Borges} et~al.}{2008}]{Borges_WENOZ2008}
{Borges} R.,  {Carmona} M.,  {Costa} B.,   {Don} W.~S.,  2008, \mn@doi [Journal
  of Computational Physics] {10.1016/j.jcp.2007.11.038}, \href
  {https://ui.adsabs.harvard.edu/abs/2008JCoPh.227.3191B} {227, 3191}

\bibitem[\protect\citeauthoryear{Boscarino, Russo  \& Semplice}{Boscarino
  et~al.}{2018}]{Boscarino2018}
Boscarino S.,  Russo G.,   Semplice M.,  2018, \mn@doi [Computers & Fluids]
  {https://doi.org/10.1016/j.compfluid.2017.10.009}, 169, 155

\bibitem[\protect\citeauthoryear{Boscheri \& Pareschi}{Boscheri \&
  Pareschi}{2021}]{Boscheri2021}
Boscheri W.,  Pareschi L.,  2021, \mn@doi [Journal of Computational Physics]
  {https://doi.org/10.1016/j.jcp.2021.110206}, 434, 110206

\bibitem[\protect\citeauthoryear{{Brecht}, {Lyon}, {Fedder}  \&
  {Hain}}{{Brecht} et~al.}{1981}]{Brecht_etal1981}
{Brecht} S.~H.,  {Lyon} J.,  {Fedder} J.~A.,   {Hain} K.,  1981, \mn@doi [\grl]
  {10.1029/GL008i004p00397}, \href
  {https://ui.adsabs.harvard.edu/abs/1981GeoRL...8..397B} {8, 397}

\bibitem[\protect\citeauthoryear{{Bucciantini} \& {Del Zanna}}{{Bucciantini} \&
  {Del Zanna}}{2006}]{Bucciantini_delZanna2006}
{Bucciantini} N.,  {Del Zanna} L.,  2006, \mn@doi [\aap]
  {10.1051/0004-6361:20054491}, \href
  {https://ui.adsabs.harvard.edu/abs/2006A&A...454..393B} {454, 393}

\bibitem[\protect\citeauthoryear{{Bucciantini} \& {Del Zanna}}{{Bucciantini} \&
  {Del Zanna}}{2013}]{Bucciantini_delZanna2013}
{Bucciantini} N.,  {Del Zanna} L.,  2013, \mn@doi [\mnras]
  {10.1093/mnras/sts005}, \href
  {http://adsabs.harvard.edu/abs/2013MNRAS.428...71B} {428, 71}

\bibitem[\protect\citeauthoryear{{Bugli}, {Del Zanna}  \&
  {Bucciantini}}{{Bugli} et~al.}{2014}]{Bugli_etal2014}
{Bugli} M.,  {Del Zanna} L.,   {Bucciantini} N.,  2014, \mn@doi [\mnras]
  {10.1093/mnrasl/slu017}, \href
  {https://ui.adsabs.harvard.edu/abs/2014MNRAS.440L..41B} {440, L41}

\bibitem[\protect\citeauthoryear{Carpenter, Kennedy, Bijl  \& et al.}{Carpenter
  et~al.}{2005}]{Carpenter2005}
Carpenter M.,  Kennedy C.,  Bijl H.,   et al. 2005, \mn@doi [J. Sci. Comput.]
  {doi.org/10.1007/BF02728987}, 25, 157

\bibitem[\protect\citeauthoryear{{Cheong}, {Pong}, {Yip}  \& {Li}}{{Cheong}
  et~al.}{2022}]{Cheong_etal2022}
{Cheong} P. C.-K.,  {Pong} D. Y.~T.,  {Yip} A. K.~L.,   {Li} T. G.~F.,  2022,
  \mn@doi [\apjs] {10.3847/1538-4365/ac6cec}, \href
  {https://ui.adsabs.harvard.edu/abs/2022ApJS..261...22C} {261, 22}

\bibitem[\protect\citeauthoryear{Conde, Gottlieb, Grant  \& et al.}{Conde
  et~al.}{2017}]{Conde2017}
Conde S.,  Gottlieb S.,  Grant Z.,   et al. 2017, \mn@doi [J. Sci. Comput.]
  {10.1007/s10915-017-0560-2}, 73, 677

\bibitem[\protect\citeauthoryear{{Dedner}, {Kemm}, {Kr{\"o}ner}, {Munz},
  {Schnitzer}  \& {Wesenberg}}{{Dedner} et~al.}{2002}]{Dedner_etal2002}
{Dedner} A.,  {Kemm} F.,  {Kr{\"o}ner} D.,  {Munz} C.-D.,  {Schnitzer} T.,
  {Wesenberg} M.,  2002, \mn@doi [Journal of Computational Physics]
  {10.1006/jcph.2001.6961}, \href
  {http://adsabs.harvard.edu/abs/2002JCoPh.175..645D} {175, 645}

\bibitem[\protect\citeauthoryear{{Del Zanna}, {Bucciantini}  \&
  {Londrillo}}{{Del Zanna} et~al.}{2003}]{DelZanna_etal2003}
{Del Zanna} L.,  {Bucciantini} N.,   {Londrillo} P.,  2003, \mn@doi [\aap]
  {10.1051/0004-6361:20021641}, \href
  {http://adsabs.harvard.edu/abs/2003A%26A...400..397D} {400, 397}

\bibitem[\protect\citeauthoryear{{Del Zanna}, {Zanotti}, {Bucciantini}  \&
  {Londrillo}}{{Del Zanna} et~al.}{2007}]{DelZanna_etal2007}
{Del Zanna} L.,  {Zanotti} O.,  {Bucciantini} N.,   {Londrillo} P.,  2007,
  \mn@doi [\aap] {10.1051/0004-6361:20077093}, \href
  {http://adsabs.harvard.edu/abs/2007A%26A...473...11D} {473, 11}

\bibitem[\protect\citeauthoryear{{Del Zanna}, {Bugli}  \& {Bucciantini}}{{Del
  Zanna} et~al.}{2014}]{DelZanna_etal_2014}
{Del Zanna} L.,  {Bugli} M.,   {Bucciantini} N.,  2014, in {Pogorelov} N.~V.,
  {Audit} E.,   {Zank} G.~P.,  eds,  Astronomical Society of the Pacific
  Conference Series Vol. 488, 8th International Conference of Numerical
  Modeling of Space Plasma Flows (ASTRONUM 2013). p.~217 (\mn@eprint {arXiv}
  {1401.3223}), \mn@doi{10.48550/arXiv.1401.3223}

\bibitem[\protect\citeauthoryear{{Del Zanna}, {Papini}, {Landi}, {Bugli}  \&
  {Bucciantini}}{{Del Zanna} et~al.}{2016}]{DelZanna_etal_2016}
{Del Zanna} L.,  {Papini} E.,  {Landi} S.,  {Bugli} M.,   {Bucciantini} N.,
  2016, \mn@doi [\mnras] {10.1093/mnras/stw1242}, \href
  {https://ui.adsabs.harvard.edu/abs/2016MNRAS.460.3753D} {460, 3753}

\bibitem[\protect\citeauthoryear{{Dionysopoulou}, {Alic}, {Palenzuela},
  {Rezzolla}  \& {Giacomazzo}}{{Dionysopoulou} et~al.}{2013}]{Dionys_etal2013}
{Dionysopoulou} K.,  {Alic} D.,  {Palenzuela} C.,  {Rezzolla} L.,
  {Giacomazzo} B.,  2013, \mn@doi [\prd] {10.1103/PhysRevD.88.044020}, \href
  {http://adsabs.harvard.edu/abs/2013PhRvD..88d4020D} {88, 044020}

\bibitem[\protect\citeauthoryear{{Dumbser} \& {Zanotti}}{{Dumbser} \&
  {Zanotti}}{2009}]{Dumbser_Zanotti2009}
{Dumbser} M.,  {Zanotti} O.,  2009, \mn@doi [Journal of Computational Physics]
  {10.1016/j.jcp.2009.06.009}, \href
  {https://ui.adsabs.harvard.edu/abs/2009JCoPh.228.6991D} {228, 6991}

\bibitem[\protect\citeauthoryear{{Evans} \& {Hawley}}{{Evans} \&
  {Hawley}}{1988}]{Evans_Hawley1988}
{Evans} C.~R.,  {Hawley} J.~F.,  1988, \mn@doi [\apj] {10.1086/166684}, \href
  {http://adsabs.harvard.edu/abs/1988ApJ...332..659E} {332, 659}

\bibitem[\protect\citeauthoryear{{Felker} \& {Stone}}{{Felker} \&
  {Stone}}{2018}]{Felker_Stone2018}
{Felker} K.~G.,  {Stone} J.~M.,  2018, \mn@doi [Journal of Computational
  Physics] {10.1016/j.jcp.2018.08.025}, \href
  {https://ui.adsabs.harvard.edu/abs/2018JCoPh.375.1365F} {375, 1365}

\bibitem[\protect\citeauthoryear{Kennedy \& Carpenter}{Kennedy \&
  Carpenter}{2019}]{Kennedy_Carpenter2019}
Kennedy C.~A.,  Carpenter M.~H.,  2019, \mn@doi [Applied Numerical Mathematics]
  {https://doi.org/10.1016/j.apnum.2018.10.007}, 136, 183

\bibitem[\protect\citeauthoryear{{Komissarov}}{{Komissarov}}{2007}]{Komissarov_2007}
{Komissarov} S.~S.,  2007, \mn@doi [\mnras] {10.1111/j.1365-2966.2007.12448.x},
  \href {http://adsabs.harvard.edu/abs/2007MNRAS.382..995K} {382, 995}

\bibitem[\protect\citeauthoryear{{Landi}, {Del Zanna}, {Papini}, {Pucci}  \&
  {Velli}}{{Landi} et~al.}{2015}]{Landi_etal2015}
{Landi} S.,  {Del Zanna} L.,  {Papini} E.,  {Pucci} F.,   {Velli} M.,  2015,
  \mn@doi [\apj] {10.1088/0004-637X/806/1/131}, \href
  {https://ui.adsabs.harvard.edu/abs/2015ApJ...806..131L} {806, 131}

\bibitem[\protect\citeauthoryear{{Londrillo} \& {del Zanna}}{{Londrillo} \&
  {del Zanna}}{2004}]{Londrillo_DelZanna2004}
{Londrillo} P.,  {del Zanna} L.,  2004, \mn@doi [Journal of Computational
  Physics] {10.1016/j.jcp.2003.09.016}, \href
  {http://adsabs.harvard.edu/abs/2004JCoPh.195...17L} {195, 17}

\bibitem[\protect\citeauthoryear{{Mattia}, {Del Zanna}, {Bugli}, {Pavan},
  {Ciolfi}, {Bodo}  \& {Mignone}}{{Mattia} et~al.}{2023}]{Mattia2023}
{Mattia} G.,  {Del Zanna} L.,  {Bugli} M.,  {Pavan} A.,  {Ciolfi} R.,  {Bodo}
  G.,   {Mignone} A.,  2023, \mn@doi [\aap] {10.1051/0004-6361/202347126},
  \href {https://ui.adsabs.harvard.edu/abs/2023A&A...679A..49M} {679, A49}

\bibitem[\protect\citeauthoryear{McCorquodale \& Colella}{McCorquodale \&
  Colella}{2011}]{Corquodale_Colella2011}
McCorquodale P.,  Colella P.,  2011, \mn@doi [Commun. Appl. Math. Comput. Sci.]
  {10.2140/camcos.2011.6.1}, 6, 1

\bibitem[\protect\citeauthoryear{{Mignone} \& {Bodo}}{{Mignone} \&
  {Bodo}}{2005}]{Mignone_Bodo_HLLC_2005}
{Mignone} A.,  {Bodo} G.,  2005, \mn@doi [\mnras]
  {10.1111/j.1365-2966.2005.09546.x}, \href
  {https://ui.adsabs.harvard.edu/abs/2005MNRAS.364..126M} {364, 126}

\bibitem[\protect\citeauthoryear{{Mignone} \& {Bodo}}{{Mignone} \&
  {Bodo}}{2006}]{Mignone_Bodo2006}
{Mignone} A.,  {Bodo} G.,  2006, \mn@doi [\mnras]
  {10.1111/j.1365-2966.2006.10162.x}, \href
  {https://ui.adsabs.harvard.edu/abs/2006MNRAS.368.1040M} {368, 1040}

\bibitem[\protect\citeauthoryear{{Mignone} \& {Del Zanna}}{{Mignone} \& {Del
  Zanna}}{2021}]{Mignone_DelZanna2021}
{Mignone} A.,  {Del Zanna} L.,  2021, \mn@doi [Journal of Computational
  Physics] {10.1016/j.jcp.2020.109748}, \href
  {https://ui.adsabs.harvard.edu/abs/2021JCoPh.42409748M} {424, 109748}

\bibitem[\protect\citeauthoryear{{Mignone}, {Plewa}  \& {Bodo}}{{Mignone}
  et~al.}{2005}]{Mignone_etal_2005}
{Mignone} A.,  {Plewa} T.,   {Bodo} G.,  2005, \mn@doi [\apjs]
  {10.1086/430905}, \href
  {https://ui.adsabs.harvard.edu/abs/2005ApJS..160..199M} {160, 199}

\bibitem[\protect\citeauthoryear{{Mignone}, {Bodo}, {Massaglia}, {Matsakos},
  {Tesileanu}, {Zanni}  \& {Ferrari}}{{Mignone}
  et~al.}{2007}]{Mignone_PLUTO2007}
{Mignone} A.,  {Bodo} G.,  {Massaglia} S.,  {Matsakos} T.,  {Tesileanu} O.,
  {Zanni} C.,   {Ferrari} A.,  2007, \mn@doi [\apjs] {10.1086/513316}, \href
  {http://adsabs.harvard.edu/abs/2007ApJS..170..228M} {170, 228}

\bibitem[\protect\citeauthoryear{{Mignone}, {Zanni}, {Tzeferacos}, {van
  Straalen}, {Colella}  \& {Bodo}}{{Mignone} et~al.}{2012}]{Mignone_PLUTO2012}
{Mignone} A.,  {Zanni} C.,  {Tzeferacos} P.,  {van Straalen} B.,  {Colella} P.,
    {Bodo} G.,  2012, \mn@doi [\apjs] {10.1088/0067-0049/198/1/7}, \href
  {http://adsabs.harvard.edu/abs/2012ApJS..198....7M} {198, 7}

\bibitem[\protect\citeauthoryear{{Mignone}, {Mattia}  \& {Bodo}}{{Mignone}
  et~al.}{2018}]{Mignone_etal_2018}
{Mignone} A.,  {Mattia} G.,   {Bodo} G.,  2018, \mn@doi [Physics of Plasmas]
  {10.1063/1.5048496}, \href
  {https://ui.adsabs.harvard.edu/abs/2018PhPl...25i2114M} {25, 092114}

\bibitem[\protect\citeauthoryear{{Mignone}, {Mattia}, {Bodo}  \& {Del
  Zanna}}{{Mignone} et~al.}{2019}]{Mignone_etal_2019}
{Mignone} A.,  {Mattia} G.,  {Bodo} G.,   {Del Zanna} L.,  2019, \mn@doi
  [\mnras] {10.1093/mnras/stz1015}, \href
  {https://ui.adsabs.harvard.edu/abs/2019MNRAS.486.4252M} {486, 4252}

\bibitem[\protect\citeauthoryear{{Miranda-Aranguren}, {Aloy}  \&
  {Rembiasz}}{{Miranda-Aranguren} et~al.}{2018}]{Miranda_etal2018}
{Miranda-Aranguren} S.,  {Aloy} M.~A.,   {Rembiasz} T.,  2018, \mn@doi [\mnras]
  {10.1093/mnras/sty419}, \href
  {http://adsabs.harvard.edu/abs/2018MNRAS.476.3837M} {476, 3837}

\bibitem[\protect\citeauthoryear{{Mizuno}}{{Mizuno}}{2013}]{Mizuno2013}
{Mizuno} Y.,  2013, \mn@doi [\apjs] {10.1088/0067-0049/205/1/7}, \href
  {http://adsabs.harvard.edu/abs/2013ApJS..205....7M} {205, 7}

\bibitem[\protect\citeauthoryear{{Munz}, {Omnes}, {Schneider},
  {Sonnendr{\"u}cker}  \& {Vo{\ss}}}{{Munz} et~al.}{2000}]{Munz_etal2000}
{Munz} C.~D.,  {Omnes} P.,  {Schneider} R.,  {Sonnendr{\"u}cker} E.,
  {Vo{\ss}} U.,  2000, \mn@doi [Journal of Computational Physics]
  {10.1006/jcph.2000.6507}, \href
  {https://ui.adsabs.harvard.edu/abs/2000JCoPh.161..484M} {161, 484}

\bibitem[\protect\citeauthoryear{{Nakamura}, {Miyoshi}, {Nonaka}  \&
  {Takahashi}}{{Nakamura} et~al.}{2023}]{Nakamura_2023}
{Nakamura} K.,  {Miyoshi} T.,  {Nonaka} C.,   {Takahashi} H.~R.,  2023, \mn@doi
  [European Physical Journal C] {10.1140/epjc/s10052-023-11343-y}, \href
  {https://ui.adsabs.harvard.edu/abs/2023EPJC...83..229N} {83, 229}

\bibitem[\protect\citeauthoryear{{Palenzuela}, {Lehner}, {Reula}  \&
  {Rezzolla}}{{Palenzuela} et~al.}{2009}]{Palenzuela_etal2009}
{Palenzuela} C.,  {Lehner} L.,  {Reula} O.,   {Rezzolla} L.,  2009, \mn@doi
  [\mnras] {10.1111/j.1365-2966.2009.14454.x}, \href
  {http://adsabs.harvard.edu/abs/2009MNRAS.394.1727P} {394, 1727}

\bibitem[\protect\citeauthoryear{Pareschi \& Russo}{Pareschi \&
  Russo}{2003}]{Pareschi_Russo2003}
Pareschi L.,  Russo G.,  2003, in Hou T.,  Tadmor E.,  eds, Hyperbolic
  Problems: Theory, Numerics, Applications: Proceedings of the Ninth
  International Conference on Hyperbolic Problems. Springer Berlin Heidelberg,
  pp 241--251, \mn@doi{10.1007/978-3-642-55711-8_21}

\bibitem[\protect\citeauthoryear{Pareschi \& Russo}{Pareschi \&
  Russo}{2005}]{Pareschi_Russo2005}
Pareschi L.,  Russo G.,  2005, \mn@doi [Journal of Scientific Computing]
  {10.1007/BF02728986}, 25, 129

\bibitem[\protect\citeauthoryear{{Pucci} \& {Velli}}{{Pucci} \&
  {Velli}}{2014}]{Pucci_Velli_2014}
{Pucci} F.,  {Velli} M.,  2014, \mn@doi [\apjl] {10.1088/2041-8205/780/2/L19},
  \href {https://ui.adsabs.harvard.edu/abs/2014ApJ...780L..19P} {780, L19}

\bibitem[\protect\citeauthoryear{{Tomei}, {Del Zanna}, {Bugli}  \&
  {Bucciantini}}{{Tomei} et~al.}{2020}]{Tomei2020}
{Tomei} N.,  {Del Zanna} L.,  {Bugli} M.,   {Bucciantini} N.,  2020, \mn@doi
  [\mnras] {10.1093/mnras/stz3146}, \href
  {https://ui.adsabs.harvard.edu/abs/2020MNRAS.491.2346T} {491, 2346}

\bibitem[\protect\citeauthoryear{{Verma}, {Teissier}, {Henze}  \&
  {M{\"u}ller}}{{Verma} et~al.}{2019}]{Verma_etal2019}
{Verma} P.~S.,  {Teissier} J.-M.,  {Henze} O.,   {M{\"u}ller} W.-C.,  2019,
  \mn@doi [\mnras] {10.1093/mnras/sty2641}, \href
  {https://ui.adsabs.harvard.edu/abs/2019MNRAS.482..416V} {482, 416}

\bibitem[\protect\citeauthoryear{{Yee}}{{Yee}}{1966}]{Yee1966}
{Yee} K.,  1966, \mn@doi [IEEE Transactions on Antennas and Propagation]
  {10.1109/TAP.1966.1138693}, \href
  {https://ui.adsabs.harvard.edu/abs/1966ITAP...14..302Y} {14, 302}

\makeatother
\end{thebibliography}



\appendix
\onecolumn

\appendix

\section{Explicit Butcher tableaux}
\label{app:butcher_tableau}
%

The following tables list the tableaux for the explicit (left), and implicit (right) coefficients for the IMEX-SSP3(4,3,3) L-stable scheme (see also \cite{Pareschi_Russo2003, Pareschi_Russo2005}, Table 6), and for the IMEX-ARK4(3)7L[2]SA$_1$ of \cite{Kennedy_Carpenter2019}.
More specifically, the explicit form of Butcher's tableaux for the IMEX-SSP3(4,3,3) consists in:

\begin{equation}\label{eq:IMEXSSP(4,3,3)}
  \begin{array}
    {c|cccc}
     0  & 0 & 0   & 0   & 0  \\
     0  & 0 & 0   & 0   & 0  \\
     1  & 0 & 1   & 0   & 0  \\
    1/2 & 0 & 1/4 & 1/4 & 0 \\
    \hline
        & 0 & 1/6 & 1/6 & 2/3
  \end{array}
  \qquad
  \begin{array}
    {c|cccc}
     \alpha  &  \alpha &   0     & 0  & 0  \\
       0     & -\alpha & \alpha  & 0  & 0  \\
       1     & 0 & 1-\alpha  & \alpha   & 0  \\
      1/2    & \beta & \eta & 1/2 -\beta -\eta -\alpha & \alpha \\
    \hline
             & 0 & 1/6 & 1/6 & 2/3 \\ \noalign{\medskip}
  \end{array}
\end{equation}

where $\alpha = 0.24169426078821$, $\beta  = 0.06042356519705$, and $\eta   = 0.12915286960590$.
Similarly, the full form of the Butcher's tableaux for the IMEX-ARK4(3)7L[2]SA$_1$ consists in:


\begin{equation}\label{eq:ARK4}
  \begin{array}
    {c|ccccccc}
     0    & 0      & 0      & 0      & 0      & 0      & 0      & 0  \\
     \Tilde{c}_2  & \Tilde{a}_{21} & 0      & 0      & 0      & 0      & 0      & 0  \\
     \Tilde{c}_3  & \Tilde{a}_{31} & \Tilde{a}_{32} & 0      & 0      & 0      & 0      & 0  \\
     \Tilde{c}_4  & \Tilde{a}_{41} & \Tilde{a}_{42} & \Tilde{a}_{43} & 0      & 0      & 0      & 0  \\
     \Tilde{c}_5  & \Tilde{a}_{51} & \Tilde{a}_{52} & \Tilde{a}_{53} & \Tilde{a}_{54} & 0      & 0      & 0  \\
     \Tilde{c}_6  & \Tilde{a}_{61} & \Tilde{a}_{62} & \Tilde{a}_{63} & \Tilde{a}_{64} & \Tilde{a}_{65} & 0      & 0  \\
     \Tilde{c}_7  & \Tilde{a}_{71} & \Tilde{a}_{72} & \Tilde{a}_{73} & \Tilde{a}_{74} & \Tilde{a}_{75} & \Tilde{a}_{76} & 0  \\
    \hline
          & 0      & 0      & \Tilde{w}_3    & \Tilde{w}_4    & \Tilde{w}_5    & \Tilde{w}_6    & \kappa 
  \end{array}
  \qquad
  \begin{array}
    {c|ccccccc}
     0    & 0      & 0      & 0      & 0      & 0      & 0      & 0  \\
     c_2  & \kappa & \kappa & 0      & 0      & 0      & 0      & 0  \\
     c_3  & a_{31} & a_{32} & \kappa & 0      & 0      & 0      & 0  \\
     c_4  & a_{41} & a_{42} & a_{43} & \kappa & 0      & 0      & 0  \\
     c_5  & a_{51} & a_{52} & a_{53} & a_{54} & \kappa & 0      & 0  \\
     c_6  & a_{61} & a_{62} & a_{63} & a_{64} & a_{65} & \kappa & 0  \\
     c_7  & a_{71} & a_{72} & a_{73} & a_{74} & a_{75} & a_{76} & \kappa  \\     \hline
          & 0      & 0      & w_3    & w_4    & w_5    & w_6    & \kappa  \\ \noalign{\medskip}
  \end{array}
\end{equation}

where $\kappa = 0.1235$. The remaining 21 explicit ($\Tilde{a}$) and the 20 implicit coefficients ($a$) are, respectively:

\begin{equation}
\begin{array}{ll}
\Tilde{a}_{21} = \DS\frac{247}{1000}  = 0.247   &  \\ \noalign{\medskip} \hline \noalign{\medskip}
\Tilde{a}_{31} = \DS\frac{247}{4000} = 0.06175  &  
\Tilde{a}_{32} = \DS\frac{2694949928731}{7487940209513}  = 0.35990537495 \\  \noalign{\medskip} \hline \noalign{\medskip}
\Tilde{a}_{41} = \DS\frac{464650059369}{8764239774964} = 0.05301658458 & 
\Tilde{a}_{42} = \DS\frac{878889893998}{2444806327765} = 0.35949264529 \\ \noalign{\medskip}
\Tilde{a}_{43} = -\DS\frac{952945855348}{12294611323341} = -0.07750922988 & \\  \noalign{\medskip} \hline \noalign{\medskip}
\Tilde{a}_{51} =  \DS\frac{476636172619}{8159180917465} = 0.05841715944 &
\Tilde{a}_{52} = -\DS\frac{1271469283451}{7793814740893} = -0.16313824817 \\ \noalign{\medskip}
\Tilde{a}_{53} = -\DS\frac{859560642026}{4356155882851} = -0.19732090979 &
\Tilde{a}_{54} =  \DS\frac{1723805262919}{4571918432560} = 0.37704199852 \\  \noalign{\medskip} \hline \noalign{\medskip}
\Tilde{a}_{61} =  \DS\frac{6338158500785}{11769362343261} = 0.5385303227 \qquad \qquad & 
\Tilde{a}_{62} = -\DS\frac{4970555480458}{10924838743837} = -0.45497746895 \\ \noalign{\medskip}
\Tilde{a}_{63} =  \DS\frac{3326578051521}{2647936831840} = 1.25629056234 &
\Tilde{a}_{64} = -\DS\frac{880713585975}{1841400956686} = -0.47828452721 \\ \noalign{\medskip}
\Tilde{a}_{65} = -\DS\frac{1428733748635}{8843423958496} = -0.16155888888 & \\  \noalign{\medskip} \hline \noalign{\medskip}
\Tilde{a}_{71} =  \DS\frac{760814592956}{3276306540349} = 0.23221715782 &
\Tilde{a}_{72} =  \DS\frac{760814592956}{3276306540349} = 0.23221715782 \\ \noalign{\medskip}
\Tilde{a}_{73} = -\DS\frac{47223648122716}{6934462133451} = - 6.80999437504 & 
\Tilde{a}_{74} =  \DS\frac{71187472546993}{9669769126921} = 7.36185855242 \\ \noalign{\medskip}
\Tilde{a}_{75} = -\DS\frac{13330509492149}{9695768672337} = -1.37487907794 &
\Tilde{a}_{76} =  \DS\frac{11565764226357}{8513123442827} = 1.35858058491  \\ \noalign{\medskip}
\end{array}
\end{equation}

and

\begin{equation}
\begin{array}{ll}
a_{31} = \DS\frac{624185399699}{4186980696204} = 0.14907768747 & 
a_{32} = \DS\frac{624185399699}{4186980696204} = 0.14907768747 \\ \noalign{\medskip} \hline \noalign{\medskip}
a_{41} = \DS\frac{1258591069120}{10082082980243} = 0.12483442871 &
a_{42} = \DS\frac{1258591069120}{10082082980243} = 0.12483442871 \\ \noalign{\medskip}
a_{43} = -\DS\frac{322722984531}{8455138723562} = -0.03816885743 \\ \noalign{\medskip} \hline \noalign{\medskip}
a_{51} = -\DS\frac{436103496990}{5971407786587} = -0.0730319403 &
a_{52} = -\DS\frac{436103496990}{5971407786587} = -0.0730319403 \\ \noalign{\medskip}
a_{53} = -\DS\frac{2689175662187}{11046760208243} = -0.24343568716 \qquad\qquad &
a_{54} = \DS\frac{4431412449334}{12995360898505} = 0.34099956776 \\ \noalign{\medskip} \hline \noalign{\medskip}
a_{61} = -\DS\frac{2207373168298}{14430576638973} = -0.15296500088 &
a_{62} = -\DS\frac{2207373168298}{14430576638973} = -0.15296500088 \\ \noalign{\medskip}
a_{63} =  \DS\frac{242511121179}{3358618340039} = 0.07220562047 &
a_{64} =  \DS\frac{3145666661981}{7780404714551} = 0.40430630248 \\ \noalign{\medskip}
a_{65} =  \DS\frac{5882073923981}{14490790706663} = 0.4059180788 \\ \noalign{\medskip} \hline \noalign{\medskip}
a_{71} =  0  &
a_{72} =  0 \\ \noalign{\medskip}
a_{73} =  \DS\frac{9164257142617}{17756377923965} = 0.51611072831 &
a_{74} = -\DS\frac{10812980402763}{74029279521829} = -0.14606356393 \\ \noalign{\medskip}
a_{75} =  \DS\frac{1335994250573}{5691609445217} = 0.23473048589 &
a_{76} =  \DS\frac{2273837961795}{8368240463276} = 0.27172234973 \\ \noalign{\medskip}
\end{array}
\end{equation}

where, for the sake of simplicity, the horizontal lines are used to highlight the different stages of the algorithm.
From Eq. \ref{eq::csumIMEXcoeff}, the explicit form of the coefficients $\Tilde{c} = c$ becomes:

\begin{equation}
\begin{array}{ll}
    \Tilde{c}_2 = \DS\frac{247}{1000} = 0.247 \qquad\qquad\qquad&
    \Tilde{c}_3 = \DS\frac{4276536705230}{10142255878289} = 0.42165537495 \\ \noalign{\medskip}
    \Tilde{c}_4 = \DS\frac{67}{200} = 0.335 &
    \Tilde{c}_5 = \DS\frac{7}{40} = 0.075 \\ \noalign{\medskip}
    \Tilde{c}_6 = \DS\frac{7}{10} = 0.7   &
    \Tilde{c}_7 = 1     \\ \noalign{\medskip}
  \end{array}
\end{equation}

The Butcher's tableaux are finally completed by the coefficients $w = \Tilde{w}$:

\begin{equation}
\begin{array}{ll}
    \Tilde{w}_3 = \DS\frac{9164257142617}{17756377923965} = 0.51611072831 \qquad\qquad&
    \Tilde{w}_4 = -\DS\frac{10812980402763}{74029279521829} = -0.14606356393 \\ \noalign{\medskip}
    \Tilde{w}_5 = \DS\frac{1335994250573}{5691609445217} = 0.23473048589 &
    \Tilde{w}_6 = \DS\frac{2273837961795}{8368240463276} = 0.27172234973  .
  \end{array}
\end{equation}


\bsp	
\label{lastpage}
\end{document}